\documentclass[prl,aps,twocolumn,showpacs,floatfix]{revtex4-1}
\usepackage{amsfonts}
\usepackage{amssymb}
\usepackage{hyperref}
\usepackage{graphicx}
\usepackage{dcolumn}
\usepackage{bm,amsmath,verbatim}
\usepackage{mathrsfs}
\usepackage{color}

\usepackage{amsmath,amssymb,amsthm}

\usepackage[T1]{fontenc}
\usepackage[latin9]{inputenc}
\usepackage{booktabs}

\hypersetup{colorlinks,
	linkcolor=blue,          citecolor=blue,        filecolor=blue,      urlcolor=blue           }

\def\be{\begin{equation}}
	\def\ee{\end{equation}}
\def\bea{\begin{eqnarray}}
	\def\eea{\end{eqnarray}}
\def\bse{\begin{subequations}}
	\def\ese{\end{subequations}}

\def\be{\begin{eqnarray}}
	\def\ee{\end{eqnarray}}

\newcommand{\ii}{\text{i}}

\begin{document}
\title{Higher-order Topological Hyperbolic Lattices}
\author{Yu-Liang Tao$^{1}$}
\author{Yong Xu$^{1,2}$}
\email{yongxuphy@tsinghua.edu.cn}
\affiliation{$^{1}$Center for Quantum Information, IIIS, Tsinghua University, Beijing 100084, People's Republic of China}
\affiliation{$^{2}$Shanghai Qi Zhi Institute, Shanghai 200232, People's Republic of China}

\begin{abstract}
A hyperbolic lattice allows for any $p$-fold rotational symmetry, in stark contrast to a two-dimensional crystalline material, where only twofold, threefold, fourfold or sixfold rotational symmetry is permitted. This unique feature motivates us to ask whether the enriched rotational symmetry in a hyperbolic lattice can lead to any new topological phases beyond a crystalline material. Here, by constructing and exploring tight-binding models in hyperbolic lattices, we theoretically demonstrate the existence of higher-order topological phases in hyperbolic lattices with eight-fold, twelve-fold, sixteen-fold or twenty-fold rotational symmetry, which is not allowed in a crystalline lattice. Since such models respect the combination of time-reversal symmetry and $p$-fold ($p=$8, 12, 16 or 20) rotational symmetry, $p$ zero-energy corner modes are protected. 
For the hyperbolic \{8,3\} lattice, we find a higher-order topological phase with a finite edge energy gap and a gapless phase.  
Our results thus open the door to studying higher-order topological phases in hyperbolic lattices.
\end{abstract}
\maketitle

\begin{figure}[htbp]
	\includegraphics[width = 1\linewidth]{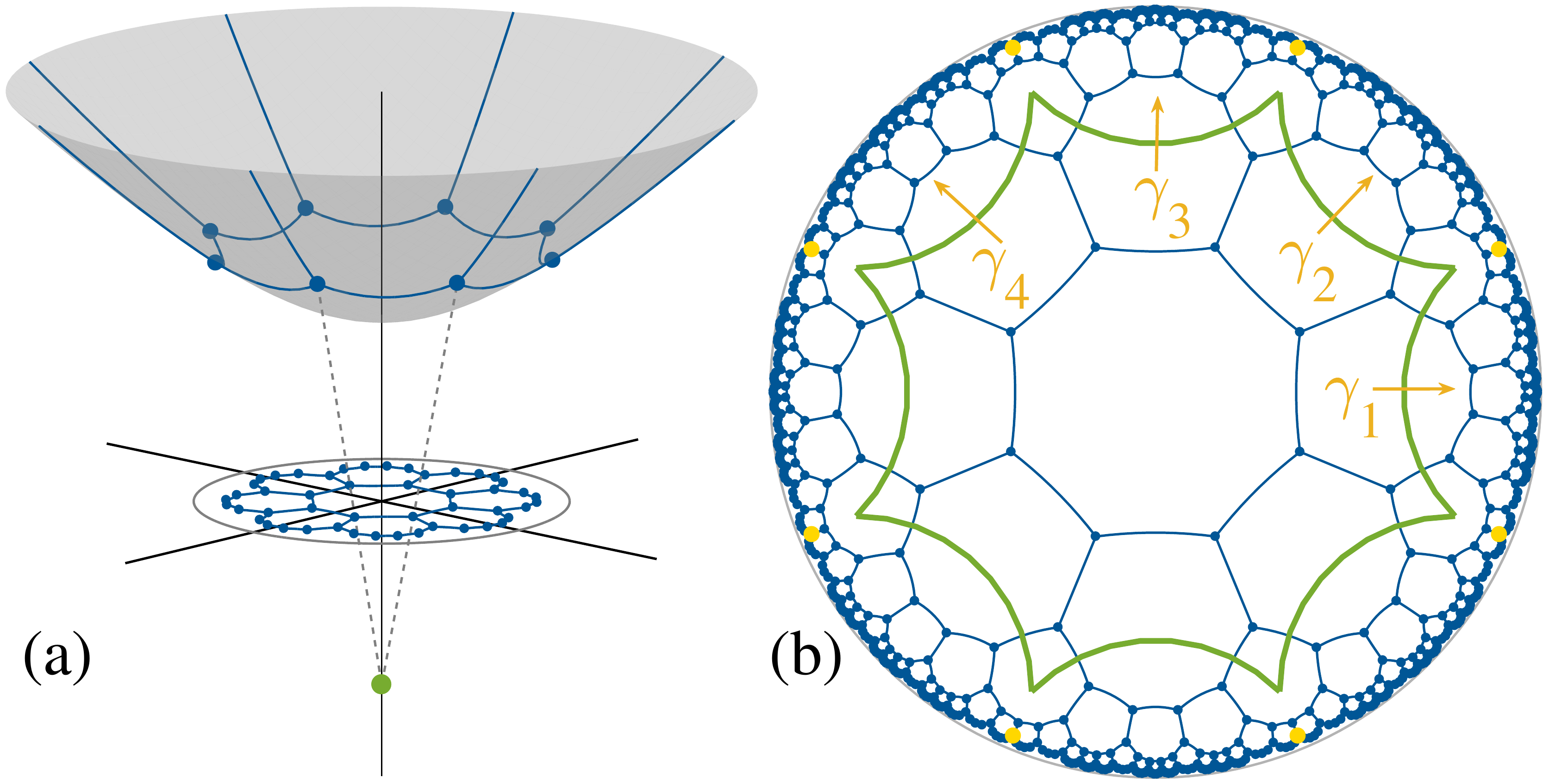}
	\caption{(a) Projection of a hyperbolic $\{8,3\}$ lattice onto a Poincar\'{e} disk. Regular
		hyperbolic octagons are used to tessellate the hyperbolic plane (described by the equation $z^2-x^2-y^2=1$) so that each lattice site is connected to
		three neighboring sites through the geodesics on the hyperbolic plane.
		The lattices on the hyperbolic plane are projected onto the unit disk on the $z=0$ plane through a line connecting points on the lattices
		to the point at $(0,0,-1)$, leading to the Poincar\'{e} disk model.
		In fact, one can achieve the tessellation through applying translational
		operations (generated by four generators $\gamma_1,\dots,\gamma_4$) to a unit cell [the region enclosed by the green curve in (b)].
		Such translational operations form a group that is non-Abelian.
		(b) The Poincar\'{e} disk model for the hyperbolic $\{8,3\}$ lattice (see Supplemental Material Sec. S-1~\cite{supplement}). 
		Solid yellow circles describe the corner modes of a higher-order topological phase on the hyperbolic lattice.
	}	
	\label{Fig1}
\end{figure}

Recently, hyperbolic lattices have been experimentally realized in circuit quantum electrodynamics~\cite{Houck2019nat} and electric circuits~\cite{Zhang2022NC,Lenggenhager2022NC,Boettcher2022arxiv}, igniting great interest in study of various properties of hyperbolic lattices~\cite{Houck2019CMP,Park2020PRL,Gorshkov2020PRA,Rayan2021scia,Matsuki2021JOP,Rouxinol2021arxiv,Rayan2022PNAS,Gorshkov2022PRL,Boettcher2022PRL,
	Thomale2022PRB,Urwyler2022arxiv,Zhou2022PRB,Mao2022arxiv,Bzdusek2022PRB,Mosseri2022PRB}, such as topological properties, hyperbolic band theory and flat band properties. Different from a hyperbolic lattice, it is a well-known fact that only twofold, threefold, fourfold or sixfold rotational symmetry is permitted in two-dimensional (2D) crystalline materials.
In other words, one can only use regular $p$-sided polygons with $p=3$, $4$ or $6$ to tessellate the 2D Euclidean plane.
However, for a hyperbolic lattice with constant negative curvature, such a restriction is lifted so that one can use regular $p$-gons for any integer $p>2$ to tessellate a hyperbolic plane [see Fig.~\ref{Fig1}(a)]. As a result, any $p$-fold rotational symmetry can be realized in a hyperbolic lattice.
This leads to a natural question of whether the enriched rotational symmetry of a hyperbolic lattice will result in any new topological phases beyond crystalline systems.

Recent generalizations of topological phases to the higher-order case provide us with an opportunity to study the effects of the enriched rotational symmetry. Different from the conventional first-order topological system, such a topological phase supports $(n-m)$-dimensional ($1<m \le n$) gapless boundary modes for an $n$-dimensional system~\cite{Taylor2017Science,Fritz2012PRL,ZhangFan2013PRL,Slager2015PRB,Brouwer2017PRL,FangChen2017PRL,Schindler2018SA,Taylor2018PRB(R),Brouwer2019PRX,Roy2019PRB,Fulga2019PRL,Yang2019PRL,Roy2020PPR,Xu2020PRL,Parameswaran2020PRL,Xu2020PPR,Xu2020PPB,AYang2020PRL,Xu2020NJP,Seradjeh2019PRB,Wang2021PRL}. For instance, a two-dimensional (2D) second-order topological insulator may support two, four or six zero-energy corner modes~\cite{Taylor2017Science,Yang2019PRL,AYang2020PRL}.
Since the number of corner modes is closely related to the crystalline symmetry of a system, one thus may wonder whether the new rotational symmetry in hyperbolic lattices can allow for the existence of new higher-order topological phases that cannot exist in a crystalline material.

In this work, we theoretically demonstrate the existence of higher-order topological phases in hyperbolic lattices
with eight-fold, twelve-fold, sixteen-fold or twenty-fold rotational symmetry by constructing and exploring tight-binding models on the lattices. Such Hamiltonians respect the combination of  time-reversal symmetry and $p$-fold ($p=$8, 12, 16 or 20) rotational symmetry, which is not allowed in a crystalline lattice.
While a quasicrystal may allow for the eight-fold or twelve-fold rotational symmetry, to the best of our knowledge, it is unclear whether sixteen-fold or twenty-fold rotational symmetry can occur there~\cite{Sunbook}. For clarify of presentation, we mainly focus on a hyperbolic $\{8,3\}$ lattice where regular $8$-gons are used to tessellate a hyperbolic plane such that each lattice site is connected to three neighboring sites [see Fig.~\ref{Fig1}(b)].
Note that for the Euclidean plane, only $\{3,6\}$, $\{4,4\}$ and $\{6,3\}$ lattices can achieve the tessellation.
For the hyperbolic $\{8,3\}$ lattice, we find a gapped and a gapless higher-order topological hyperbolic phase by numerically computing the quadrupole moment and energy band properties of a four-dimensional (4D) momentum-space Hamiltonian based on the hyperbolic band theory~\cite{Rayan2021scia}.
We further study a Hamiltonian on a hyperbolic lattice without translational symmetry and find the 
existence of eight-fold degenerate zero-energy modes localized at eight corners in a phase with a finite edge energy gap. 
The topology of this phase is characterized by the corner charge. Interestingly, there also appears a gapless phase with vanishing edge energy gap where
states near zero energy are mainly localized on the boundary of the hyperbolic lattice. 
In addition, the real space results also show a reentrant gapped topological phase with zero-energy corner modes,
which may arise from finite-size effects.
Finally, we show the existence of twelve, sixteen or twenty zero-energy corner modes in a hyperbolic lattice with the corresponding rotational symmetry.

\emph{Model with hyperbolic translational symmetry}.---To demonstrate the existence of higher-order topological phases in hyperbolic lattices, we will construct two types of
tight-binding models in a hyperbolic lattice described by the Poincar\'{e} disk model [see Fig.~\ref{Fig1}(b)].
We start by constructing the first tight-binding model in a hyperbolic ${\{8,3\} }$ lattice with hyperbolic translational symmetry by following Ref.~\cite{Urwyler2022arxiv}.
We first construct the onsite and hopping term inside the first unit cell [the region enclosed by the green curve in Fig.~\ref{Fig1}(b)],
${H}_0=\sum_{\alpha,\beta}[\sum_{i} m |1 i\alpha \rangle [\tau_z\sigma_0]_{\alpha \beta} \langle  1 i \beta| 
+ \sum_{\left\langle i,j \right\rangle } |1 i \alpha \rangle [T(\theta_{ij})]_{\alpha \beta} \langle 1 j \beta|]$,
where $|r i \alpha\rangle$ denotes the state of the $\alpha$th degree of freedom at the site $i$ in the $r$th unit cell in the Poincar\'{e} disk. 
At each site, there are four degrees of freedom, and $\left\{\tau_\nu\right\}$ and $\left\{\sigma_\nu\right\}$ with $\nu=x,y,z$ are two sets 
of Pauli matrices that act on these degrees of freedom. In $H_0$, the first term $m\tau_z\sigma_0$ describes the on-site energy, and
the second term depicts the hopping between two neighboring connected sites inside the unit cell with the hopping matrix
$T(\theta_{ij})=[t_0{\tau _z}{\sigma _0}-{\ii} {t_1}( \cos \theta _{ij}\tau _x\sigma _x + \sin \theta _{ij}\tau_x\sigma_y )
+ g\cos(p\theta _{ij}/2)\tau _y \sigma _0]/2$ with $p=8$, where $\theta _{ij}$ is the polar angle of the vector from the site $j$ to site $i$ in the first unit cell.
In the following, we will set the system parameters $t_0=t_1=1$ as the units of energy.

For the hopping term between the sites in the first unit cell and
the sites in the neighboring four unit cells described by the set $S_1$, we define it as
$H_1=\sum_{r\in S_1}\sum_{\alpha,\beta} \sum_{\left\langle i,j \right\rangle } |r i\alpha \rangle [T(\tilde{\theta}_{(ri),(1j)})]_{\alpha \beta} \langle 1 j \beta|+\text{H.c.} $.
To ensure that the system has $C_pT$ symmetry (here $p=8$) where 
$C_p$ is the $p$-fold rotational operator and $T$ is the time-reversal operator, we have to consider a modified $\tilde{\theta}_{(ri),(1j)}$ (see Supplemental
Material Sec. S-2). The entire Hamiltonian can be generated by applying 
translational operations (generated by generators $\gamma_1,\dots,\gamma_{4}$).

When $g=0$, the system respects in-plane mirror symmetry $M_1=\tau_z \sigma_z$, time-reversal symmetry ${T}=i\sigma_y \kappa$ where $\kappa$ is the complex conjugate operator, chiral symmetry $\Gamma=\tau_x\sigma_z$, and thus particle-hole symmetry $\Xi=\tau_x \sigma_x \kappa$. Owing to the eight-fold rotational symmetry about the $z$ axis preserved by the hyperbolic lattice, the Hamiltonian also respects the eight-fold rotational symmetry
${C}_p=\tau_0 e^{-i\frac{\pi}{p}\sigma_z}R_p$ where $R_p|i \alpha \rangle\equiv |g_p(i) \alpha \rangle$ with $g_p(i)$ rotating the lattice site $i$ by an angle $2\pi/p$ about the $z$ axis
(here $p=8$).
With these internal symmetry, the system belongs to the DIII class corresponding to a $\mathbb{Z}_2$ topological insulator whose nontrivial phase exhibits helical edge modes. To generate a higher-order phase with corner modes, we
add the term $g\cos(4\theta _{ij})\tau _y \sigma _0$ to break the time-reversal symmetry so as to open the gap of the helical modes at a boundary; this term thus acts as an edge mass term. As this term changes its sign once $\theta _{ij}$ increases by $\pi/4$, a corner state may arise at the location where the mass flips its sign. Since the change of sign occurs eight times, total eight corner modes may appear.
While the term also breaks the $C_8$ symmetry, the combination of $T$ and $C_8$ symmetry is still conserved.
The symmetry ensures that the number of corner modes must be an integer multiple of eight given the fact that if there is a zero-energy corner mode $|\psi_c\rangle$ mainly
localized at ${\bf r}$, then
$C_8T|\psi_c\rangle$ is also a zero-energy corner mode mainly localized at $g_8({\bf r})$.

We now employ the twisted boundary conditions to construct the momentum-space Bloch Hamiltonian based on the hyperbolic band theory~\cite{Rayan2021scia,Rayan2022PNAS,Mao2022arxiv}. In the Hamiltonian, the hopping between two sites in two different unit cells which are connected by a translation operator $\gamma_j$ should carry an extra phase term $e^{-\text{i}k_j}$. We thus obtain a $64\times 64$ Hamiltonian $H({\bf k})=H(k_1,k_2,k_3,k_4)$ in a four-dimensional Brillouin zone with $k_j\in[0,2\pi]$ for $j=1,\dots,4$ (see Supplemental
Material Sec. S-2) .

While time-reversal symmetry is broken in the Hamiltonian $H$, chiral symmetry is still preserved so that $H({\bf k})$ respects chiral symmetry.
In light of the fact that the quadrupole moment~\cite{Cho2019PRB,Hughes2019PRB} is protected to be quantized by chiral symmetry~\cite{Xu2021PRB,Shen2020PRL},
we can utilize the quadrupole moment to characterize the topological property of $H({\bf k})$ spanned by two of the four momenta.
Specifically, one can regard $H({\bf k})$ spanned by $k_i$ and $k_j$ ($i,j\in {1,\dots,4}$ and $i\neq j$) with the other two momenta
$k_{\bar{i}}$ and $k_{\bar{j}}$ fixed as the momentum-space version of a Hamiltonian $H_s$ in an $L\times L$ square lattice.
The quadrupole moment for the occupied states is defined as~\cite{Cho2019PRB,Hughes2019PRB,Xu2021PRB,Shen2020PRL}
\begin{equation}
	\label{Qxyij}
	Q_{ij}(k_{\bar{i}},k_{\bar{j}})=[\frac{1}{2\pi} \mathrm{Im}\log \det (U_o^\dag \hat{D}U_o )-Q_0 ]\ \text{mod}\ 1,
\end{equation}
where $U_{o}=\left(|\psi_1 \rangle,|\psi_2 \rangle,\cdots,|\psi_{n_c} \rangle\right)$ with $|\psi_j \rangle$ being the $j$th occupied eigenstate of
$H_s$ (one of $n_c=32L^2$ occupied states), and $\hat{D}=\mathrm{diag}\left\{e^{2\pi \text{i} x_jy_j/L^2}\right\}_{j=1}^{64L^2}$
with $(x_j,y_j)$ denoting the square position of the $j$th lattice site.
Here, $Q_0$ is the contribution from the background positive charge distribution.

To distinguish between an insulating phase and a semimetal phase, we calculate the average quadrupole moment over all fixed momenta $(k_{\bar{i}},k_{\bar{j}})$
\begin{equation}
	\label{Qxy}
	\overline{Q}_{ij}=\frac{1}{(2\pi)^2}\int dk_{\bar{i}} dk_{\bar{j}} Q_{ij}(k_{\bar{i}},k_{\bar{j}}).
\end{equation}
The system respects $C_8M_1$ symmetry, that is,
$
	U_{C_8M_1}H(k_1,k_2,k_3,k_4)(U_{C_8M_1})^{-1}=H(-k_4,k_1,k_2,k_3)
$.
It follows that $\overline{Q}_{ij}$ should satisfy the following relations (see Supplemental Material Sec. S-3 for proof):
\begin{align}
	\label{Qxyrelation}
	\overline{Q}_{12}=\overline{Q}_{23}=&\overline{Q}_{34}=\overline{Q}_{41} \\ \nonumber
	\text{and}\qquad &\overline{Q}_{13}=\overline{Q}_{24}.
\end{align}
Note that a similar relation for the Chern number has been derived in Ref.~\cite{Urwyler2022arxiv}.
We find that $\overline{Q}_{13}$ is always equal to zero and thus use $\overline{Q}_{12}$ to characterize the topological property.

\begin{figure}[t]
	\includegraphics[width = 1\linewidth]{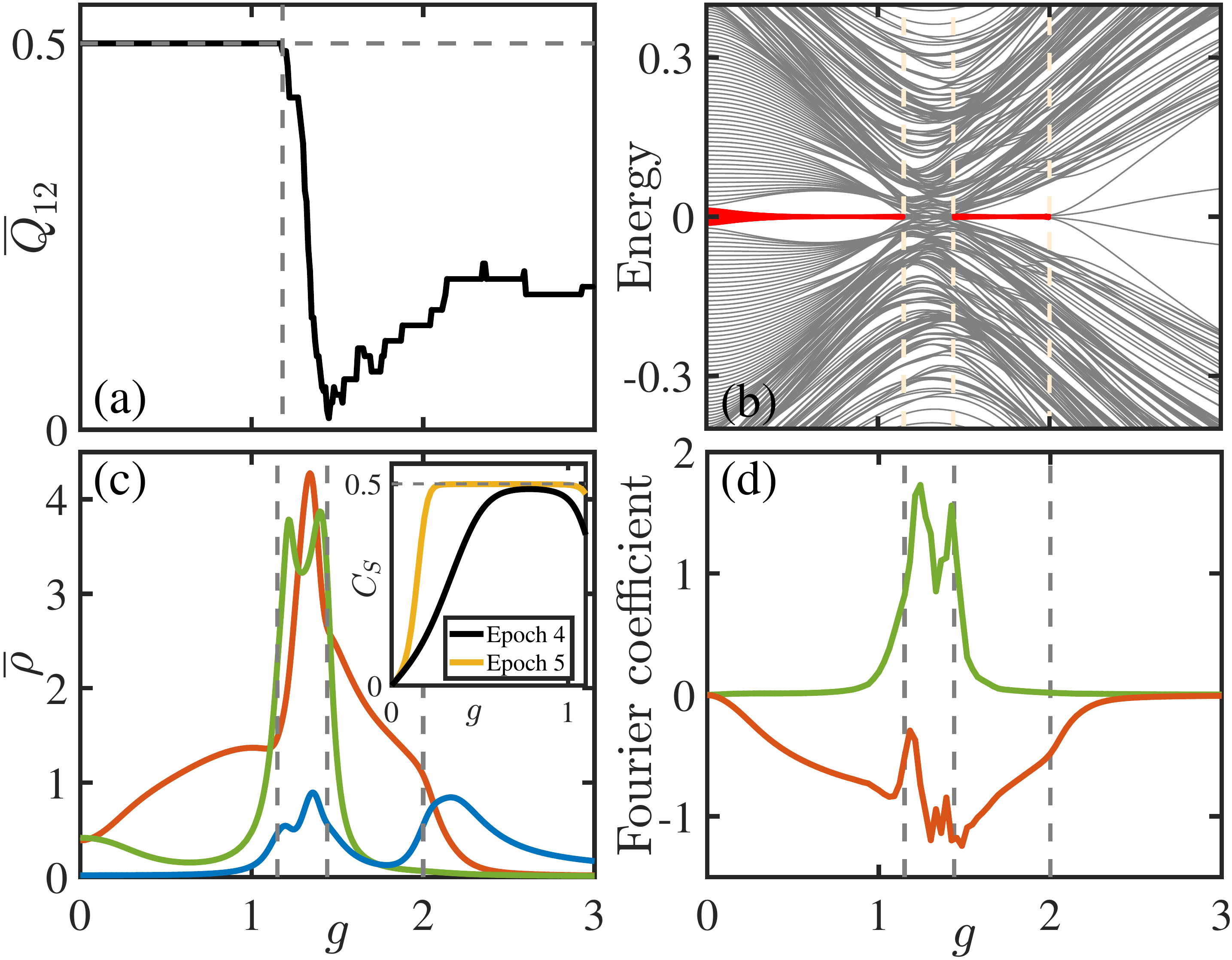}
	\caption{Topological and band properties of a hyperbolic $\{8,3\}$ lattice. (a) $\overline{Q}_{12}$ as a function of $g$.  
	(b) The energy spectrum of the tight-binding Hamiltonian in Eq.~(\ref{TB}) under open boundary conditions with respect to $g$. 
	Red lines highlight the zero-energy corner modes. (c) Zero-energy average DOS defined as $\overline{\rho}=(1/N)\sum_{{\bf r} \in S}{\rho(E=0,{\bf r})}$ 
	at the position near a corner (red line), the center of an edge (green line) and in the bulk (blue line). 
	For the red (green) line, $S$ is a set consisting of eleven (twelve) sites on the boundary in the vicinity of a corner (the center of an edge).
	For the bulk one, $S$ contains sixteen sites located in the central unit cell. $N$ denotes the number of elements in $S$.
	Inset: the corner charge versus $g$ at epoch 4 (black line) and 5 (yellow line). 
 (d) The Fourier coefficient $d_{-}$ (red line) and $d_+$ (green line) of the local DOS $\rho_-$ and $\rho_+$, respectively. (see the text on how to evaluate the coefficient). 
 In (b)-(d), the system has 768 sites at epoch 4, and four distinct phases are separated by three dashed vertical lines.
   Here, $m=0.8$. }	
	\label{Fig2}
\end{figure}

The average quadrupole moment illustrates a sharp decline from a quantized value of $0.5$ to a nonzero fractional value as $g$ increases as shown in Fig.~\ref{Fig2}(a),
indicating the presence of two distinct phases. One phase is a higher-order topological hyperbolic insulator with $\overline{Q}_{12}=0.5$. The other one is
a higher-order topological hyperbolic semimetal with vanishing energy gap in the 4D momentum space~\cite{footnote}. In fact, there are several degenerate nodes
in momentum space. In the Brillouin zone spanned by $(k_3,k_4)$, a part has the quadrupole moment $Q_{12}$ of $0.5$ and the other part has zero $Q_{12}$, leading to a fractional value of the average quadrupole moment (see Supplemental Material Sec. S-4). The existence of the gapless phase is in stark contrast to the higher-order phase on Euclidean square $\{4,4\}$ lattices 
where the system is always gapped as we increase the term that breaks the time-reversal symmetry~\cite{Taylor2017Science,Schindler2018SA}.

\emph{Model without hyperbolic translational symmetry}.---For the model with translational symmetry, while the hyperbolic band theory 
predicts the existence of higher-order topological phases, we find that its energy spectrum in real space changes dramatically for
different system sizes possibly due to large boundary effects (see Supplemental Material Sec. S-5). We therefore construct another tight-binding model 
Hamiltonian in a hyperbolic ${\{p,q\} }$ lattice as
\begin{equation}\label{TB}
	{H}=\sum_{\alpha,\beta}[\sum_{i} m |i\alpha \rangle [\tau_z\sigma_0]_{\alpha \beta} \langle i \beta| +
	\sum_{\left\langle i,j \right\rangle } |i\alpha \rangle [T(\theta_{ij})]_{\alpha \beta} \langle j \beta| ],
\end{equation}
where 
$|i \alpha\rangle$ denotes the state of the $\alpha$th degree of freedom at the site $i$ in the disk, and
$\theta_{ij}$ is the polar angle of the vector from the site $j$ to the site $i$, rather than a modified one.
This model respects the $C_pT$ symmetry.
Note that $p$ should be an integer multiple of 4 to ensure that the Hamiltonian is Hermitian. 

To illustrate the existence of zero-energy corner modes, we calculate the eigenenergies and eigenstates of the Hamiltonian in Eq.~(\ref{TB}) in real space under open boundary conditions.
We further compute the local density of states (DOS) defined as
$	\rho(E,\textbf{r})=\sum_{i,j} \delta (E-E_i)|\Psi_{i,j}(\textbf{r})|^2$,
where  $\Psi_{i,j}(\textbf{r})$ is the $j$th component of the $i$th eigenstate at site $\textbf{r}$ with the eigenenergy $E_i$.

\begin{figure}[t]
	\includegraphics[width = 1\linewidth]{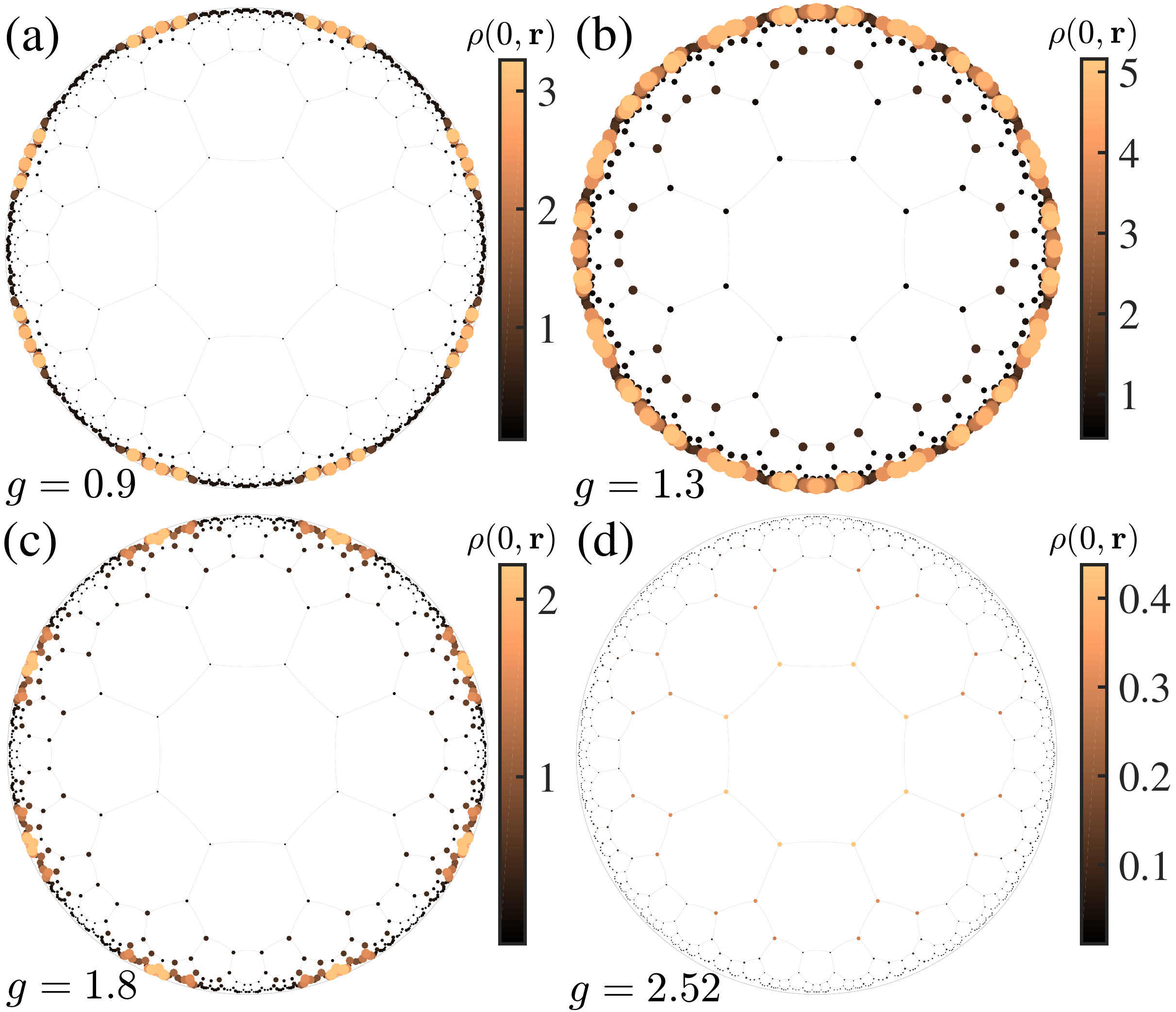}
	\caption{The local DOS $\rho(E,\textbf{r})$ at zero energy. Each figure corresponds to a typical local DOS for each phase shown in Fig.~\ref{Fig2}(b).}	
	\label{Fig3}
\end{figure}

The energy spectrum in Fig.~\ref{Fig2}(b) shows the existence of eight-fold degenerate zero-energy states when $0<g<1.15$. These modes are mainly
localized near the corner positions on the boundary at the polar angle $\theta_c=\pi/8+n\pi/4$ with $n=0,1,\dots,7$ as shown by the local DOS in 
Fig.~\ref{Fig3}(a). Such a feature is also revealed by the average local DOS near a corner and the center of an edge (e.g., $\theta=0$) 
(here an edge refers to the collection of the boundary sites between two nearest neighboring corners).
When $g=0$, the system is in the first-order topological phase, and thus the gapless edge states are almost equally distributed on the edge and corner sites.  
As we increase $g$, the energy gap at the edge is opened, leading to the appearance of corner modes, reflected by the increase (decrease) of the 
average local DOS at the corner (the center of an edge) [Fig.~\ref{Fig2}(c)].
In this regime, the average local DOS in the bulk almost vanishes. 

To further characterize its topology, we calculate the corner charge $C_S$ (see Supplemental Material S-6 for its definition) and find that it approaches the quantized value of 
$0.5$ as shown in the inset of Fig.~\ref{Fig2}(c). In the Supplemental Material S-7, we show that with increasing the system size, while 
the mini-gap (determined by the first nonzero eigenenergy) declines, the edge energy gap (determined by the energy where the degeneracy suddenly drops
from the eight-fold to the double one) remains almost unchanged.
More interestingly, for a larger system, the corner charge becomes closer to $0.5$, indicating a better topological behavior,
which arises due to a significant drop of the energy splitting of the zero-energy states.
These results strongly suggest the existence of the topological phase in the thermodynamic limit.
See Supplemental Material S-7 for finite-size analysis of the energy gap and local DOS.

As we further increase $g$, the energy spectrum in Fig.~\ref{Fig2}(b) becomes continuous near zero ($1.15 < g < 1.44$), leading to a gapless phase (vanishing edge energy gap)
with finite local DOS in the bulk [Fig.~\ref{Fig2}(c)]. 
The local DOS at zero energy in Fig.~\ref{Fig3}(b) exhibits the distribution mainly localized on boundaries, implying that the 
gapless modes are mainly comprised of boundary modes including the states localized at corner positions and other positions at the boundary. 
Such a fact is also revealed by a significant rise of the average local DOS near a corner and the center of an edge when we enter into this regime [see Fig.~\ref{Fig2}(c)].

To show how the states near zero energy affects the local DOS when the system enters into the gapless phase,
we approximate the density of an eigenstate by $n_{i}(\theta)=\sum_j|\Psi_{i,j}|^2 \approx a+b\cos (8\theta)$, 
which is the expansion of the density
in a Fourier series up to the first order. Here, we write the density as a function of the polar angle $\theta$ in the Poincar\'{e} disk 
for each site on the boundary, and $n_{i}(\theta+\pi/4)=n_{i}(\theta)$ due to the $C_8T$ symmetry. 
For this approximate density, 
if $b<0$, then it takes a maximum value at the corner position, i.e., $\theta=\theta_c$; 
if $b>0$, then the maximum occurs at the center of two neighboring corners, i.e., $\theta=\theta_e=n\pi/4$ with $n$ being an integer.
For each state, we calculate the Fourier coefficient $b$ and then 
use the states with $b<0$ ($b>0$) to calculate the corresponding local DOS $\rho_{-}$ ($\rho_{+}$).
We then expand $\rho_{\pm}$ as $c_{\pm}+d_{\pm}\cos (8\theta)$ and plot the Fourier coefficient $d_{\pm}$ in Fig.~\ref{Fig2}(d). 
In the region $0<g <1.15$, $d_-$ ($d_+$) is finite and negative (vanishes), consistent with our previous results
that the eight zero-energy modes are mainly localized at the position near $\theta_c$.
When the system enters into the gapless regime, we see a sharp rise of $d_+$, suggesting
the appearance of states near zero energy with large occupation close to the center of an edge 
(see Supplemental Material Sec. S-8 for more discussion).

In Fig.~\ref{Fig2}(b), we also see that with the further increase of $g$ until 2, the energy spectrum becomes gapped again with eight zero-energy states separated from the other states.
The phase arises because of the overlap between edge wave functions, leading to the energy gap opening of the edge states.
Such an overlap is reflected by the sudden drop of the proportion of edge states on the boundary with respect to $g$ [see Fig.~S9(a) in Supplemental Material].
In other words, the finite-size effect opens the gap of the edge wave functions, leaving the corner modes at zero energy [also evidenced by the zero-energy 
local DOS in Fig.~\ref{Fig3}(c)]. The finite-size effect will be reduced by increasing the system size so that the gapless regime 
becomes smaller (see Supplemental Material Sec. S-7).
In addition,
when $g>2$, zero-energy corner modes bifurcate into four branches away from zero energy, and no corner states are observed
in this phase [Fig.~\ref{Fig3}(d)] (see Supplemental Material Sec. S-7).

\begin{figure}[t]
	\includegraphics[width = 1\linewidth]{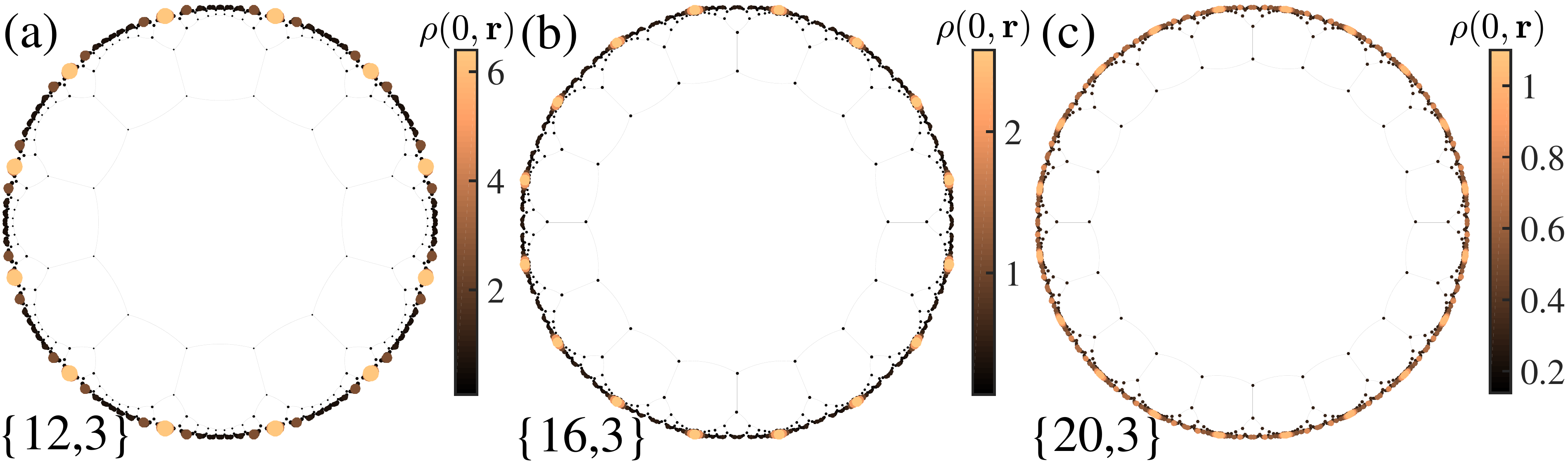}
	\caption{The local DOS at zero energy for hyperbolic (a) $\{12,3\}$ lattice with $m=0.9$ and $g=0.8$, (b) $\{16,3\}$ lattice with $m=0.975$ and $g=0.57$, and (c) $\{20,3\}$ lattice with $m=1$ and $g=0.6$. }	
	\label{Fig4}
\end{figure}

\emph{Higher-order topological phases with twelve, sixteen or twenty corner modes}.---We now proceed to 
study the higher-order topological phase in a hyperbolic lattice respecting twelve-fold, sixteen-fold or twenty-fold rotational symmetry
by constructing the Hamiltonian in hyperbolic $\{12,3\}$, $\{16,3\}$, or $\{20,3\}$ lattices.
Such Hamiltonians respect chiral symmetry and the corresponding $C_pT$ symmetry with $p=12$, $16$, or $20$. 
Figure~\ref{Fig4} illustrates the local DOS at zero energy in the Poincar\'{e} disk, indicating the existence of twelve, sixteen and twenty 
corner modes, respectively (see the energy spectrum in Supplemental Material S-9). All these phases arise from the allowed rotational symmetry of a hyperbolic lattice, and thus
cannot exist in a crystalline material. Even in a quasicrystal, it is unclear whether sixteen-fold or twenty-fold rotational symmetry can occur, 
to the best of our knowledge~\cite{Sunbook}.

In summary, we have theoretically predicted higher-order topological phases in hyperbolic lattices with eight-fold, twelve-fold, sixteen-fold or twenty-fold rotational symmetry,
which have eight, twelve, sixteen and twenty corner modes, respectively. 
Such phases are not allowed in a crystalline material.
For the hyperbolic $\{8,3\}$ lattices, we identify a higher-order topological phase with a finite edge energy gap
and a gapless phase.
Given that hyperbolic lattices have been experimentally realized in circuit quantum electrodynamics~\cite{Houck2019nat} and electric circuits~\cite{Zhang2022NC,Lenggenhager2022NC,Boettcher2022arxiv}, higher-order topological phases in hyperbolic lattices may be observed in these systems.
In fact, higher-order topological phases in square lattices have been observed in phononic~\cite{Huber2018Nat}, microwave~\cite{Bahl2018Nat}, electric circuit~\cite{Thomale2018NP}, and photonic systems~\cite{Hafezi2019NP}. Our work may also inspire the interest of studying higher-order topological phases in hyperbolic lattices in quantum simulators, such as cold atom systems~\cite{Browaeys2016Science,Lukin2016Science}. 

\begin{acknowledgments}
	We thank J.-H. Wang for helpful discussions.
	This work is supported by the National Natural Science Foundation of China (Grant No. 11974201) and Tsinghua University Dushi Program.
\end{acknowledgments}

\begin{widetext}
	\setcounter{equation}{0} \setcounter{figure}{0} \setcounter{table}{0} %
	\renewcommand{\theequation}{S\arabic{equation}} \renewcommand{\thefigure}{S%
		\arabic{figure}} \renewcommand{\bibnumfmt}[1]{[S#1]} 
	\renewcommand{\citenumfont}[1]{S#1}
	
	In the Supplemental Material, we will give a pedagogical introduction to how the hyperbolic lattices on the Poincar\'{e} disk 
		are generated based on circular inversion in Section S-1,
		show how the modified angle is defined and present the momentum space Hamiltonian in Section S-2,
		prove that $\overline{Q}_{12}=\overline{Q}_{23}=\overline{Q}_{34}=\overline{Q}_{41}$ in Section S-3,
		provide more detailed discussion on the band and topological properties of the higher-order topological hyperbolic semimetal phase in 
		the 4D momentum space in Section S-4,
		provide the energy spectrum and local DOS for the model with translational symmetry calculated in a geometry with open boundaries in Section S-5,
		utilize the corner charge to characterize the gapped topological phases in Section S-6,
		show that the higher-order topological phase continues to exist in a much larger system even though the mini-gap becomes very small
		and how the reentrant gapped phase and the branching arise in Section S-7,
		provide the density profile of the states in the vicinity of zero energy in the gapless phase in Section S-8,
		present the energy spectra of the hyperbolic \{12,3\}, \{16,3\} and \{20,3\} lattices with respect to the system parameter $g$ in Section S-9, and finally
		present the effect of weak disorder on topological phases in Section S-10.
	
		\section{S-1. Generation of hyperbolic lattices based on circular inversion}
		In this section, we will follow Ref.~\cite{HartshornebookS} to give a pedagogical introduction to how the hyperbolic lattices on the Poincar\'{e} disk 
		are generated based on circular inversion.
		
		Hyperbolic lattices can be described by the Poincar\'{e} disk model, where points are written as $z=x+\text{i}y=re^{\text{i}\theta}$. The metric on this disk is 
		\begin{equation}
			\label{metric}
			\text{d}s^2=(2\kappa_0)^2\frac{\text{d}x^2+\text{d}y^2}{(1-r^2)^2},
		\end{equation}
		where $\kappa_0$ is the curvature radius, which is equal to the radius $R$ of the disk (here we set $R=1$). 
		According to this metric, the geodesic line between points $A$ and $B$ is the circular arc of an inversion circle. 
		
		Consider the outmost cirlce of the disk with center $O$ [see Fig.~\ref{circle}(a)].
		We now show how to obtain another circle $O^\prime$ determined by the points $A$, $B$ and their inverses with respect to the circle $O$.
		The inverse of $A$ (marked as $A^\prime$) (similarly for $B$) lies on the ray from $O$ to $A$ satisfying
		\begin{equation}
			\label{inverion}
			OA\cdot OA'=R^2.
		\end{equation}
		These four points lie on the inversion circle $O^\prime$ with center $O^\prime$ and redius $R^\prime$. The geodesic line between $A$ and $B$ on the disk is the circular arc of the inversion circle $O^\prime$.
		\begin{figure}[htbp]
			\includegraphics[width = 0.7\linewidth]{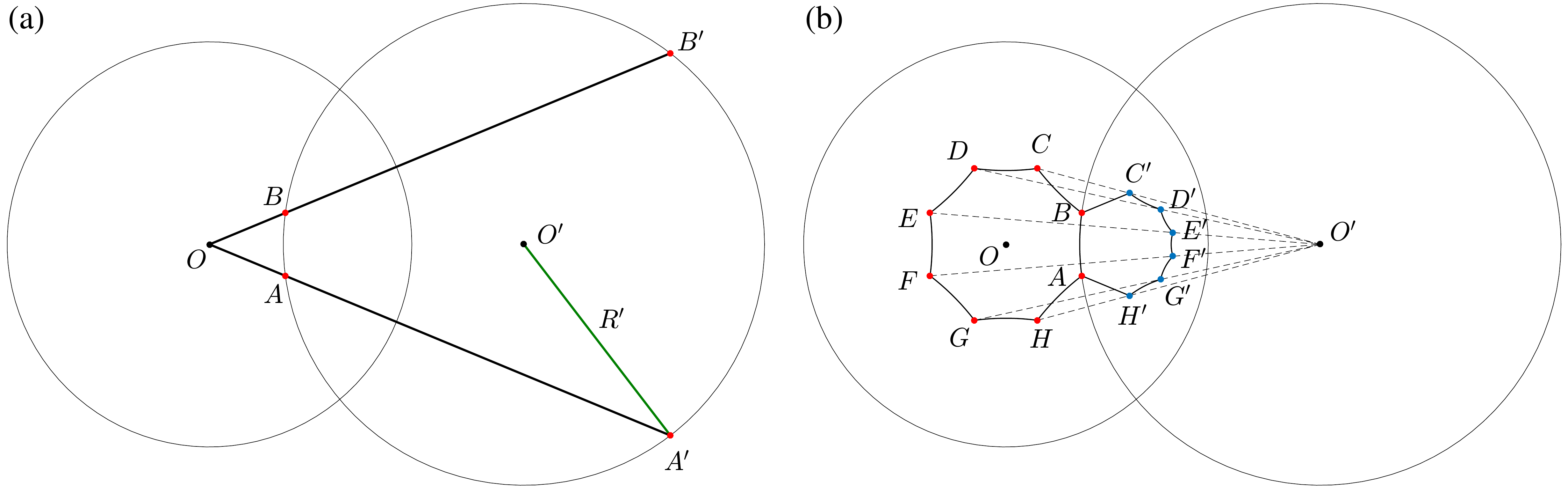}
			\caption{(a) Schematic of how to determine an inversion circle $O^\prime$ with center $O'$ and radius $R'$. 
				Points $A^\prime$ and $B^\prime$ are the inverses of $A$ and $B$ with respect to the circle $O$. 
				The inversion circle $O^\prime$ is the circle on which the four points $A$, $B$, $A^\prime$ and $B^\prime$ lie. 
				(b) Schematic of how the vertices of a new octagon adjacent to an edge $AB$ are generated.
				The vertices $C^\prime, \dots, H^\prime$ of the new polygon correspond to the inversion points of $C,\dots,H$ with respect to the circle $O^\prime$.
				Two neighboring points are connected by the geodesic lines, which is the circular arc of an inversion circle determined by these two points.}	
			\label{circle}
		\end{figure}
		
		We are now in a position to present the process of constructing a hyperbolic $\{p,q\}$ lattice with inversion circles. First, we plot the central polygon with $p$ vertices and connect
		two neighboring ones through geodesic lines, 
		which is the lattice at epoch 1 [see Fig.~\ref{epoch}(a)]. 
		These $p$ vertices are determined by $z_j=r_0e^{\text{i}(2\pi j/p+\delta)}$ where $j=1,\dots,p$, $\delta$ is an arbitrary value, and $r_0$ is the distance from the center $O$ to a vertex 
		determined by
		\begin{equation}
			\label{r0}
			r_0=\sqrt{\frac{\cos\left(\frac{\pi}{p}+\frac{\pi}{q}\right)}{\cos\left(\frac{\pi}{p}-\frac{\pi}{q}\right)}}.
		\end{equation}
		Second, we show how to generate a new polygon adjacent to an edge (e.g., $AB$) of the first one.
		The vertices $C',\dots,H'$ of this polygon are the inverses of $C,\dots, H$ with respect to the circle $O^\prime$ [see Fig.~\ref{circle}(b)].
		Similarly, one can obtain the vertices adjacent to other edges, ending up with a lattice at epoch 2 [see Fig.~\ref{epoch}(b)].
		In Fig.~\ref{epoch}(c) and (d), we also plot the hyperbolic lattices at epoch 3 and 4, respectively. 
		\begin{figure}[htbp]
			\includegraphics[width = 0.7\linewidth]{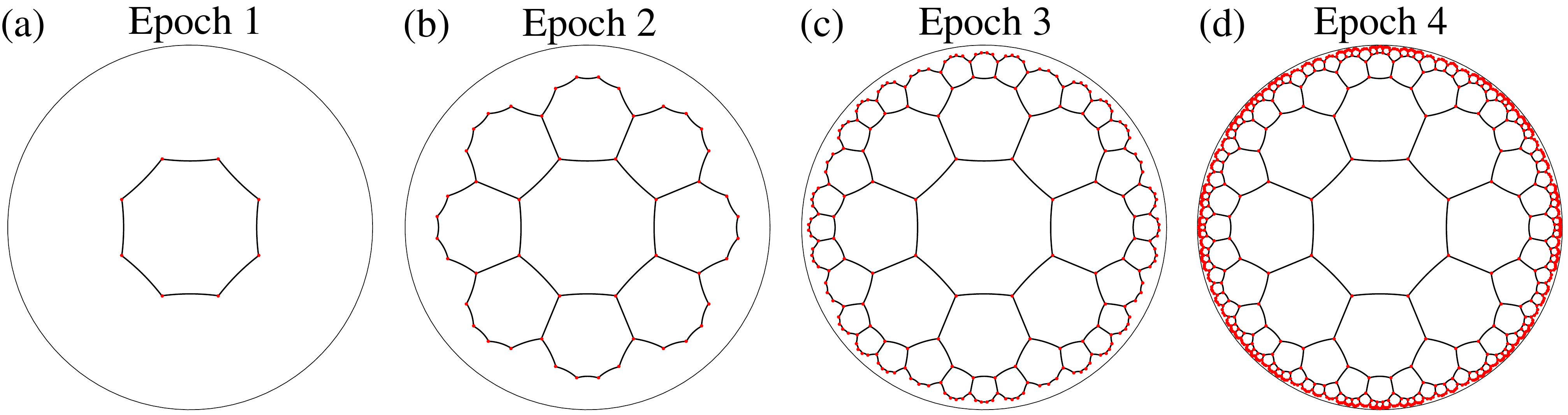}
			\caption{The hyperbolic \{8,3\} lattice with $\delta=\pi/8$ at (a) epoch 1, (b) epoch 2, (c) epoch 3, and (d) epoch 4. Two neighboring sites are connected by geodesic lines.
				The numbers of vertices in these figures are 8, 48, 200 and 768, respectively.}	
			\label{epoch}
		\end{figure}
		
		\section{S-2. Model with hyperbolic translational symmetry and the momentum space Hamiltonian}
		\begin{figure}[htbp]
			\includegraphics[width = 0.3\linewidth]{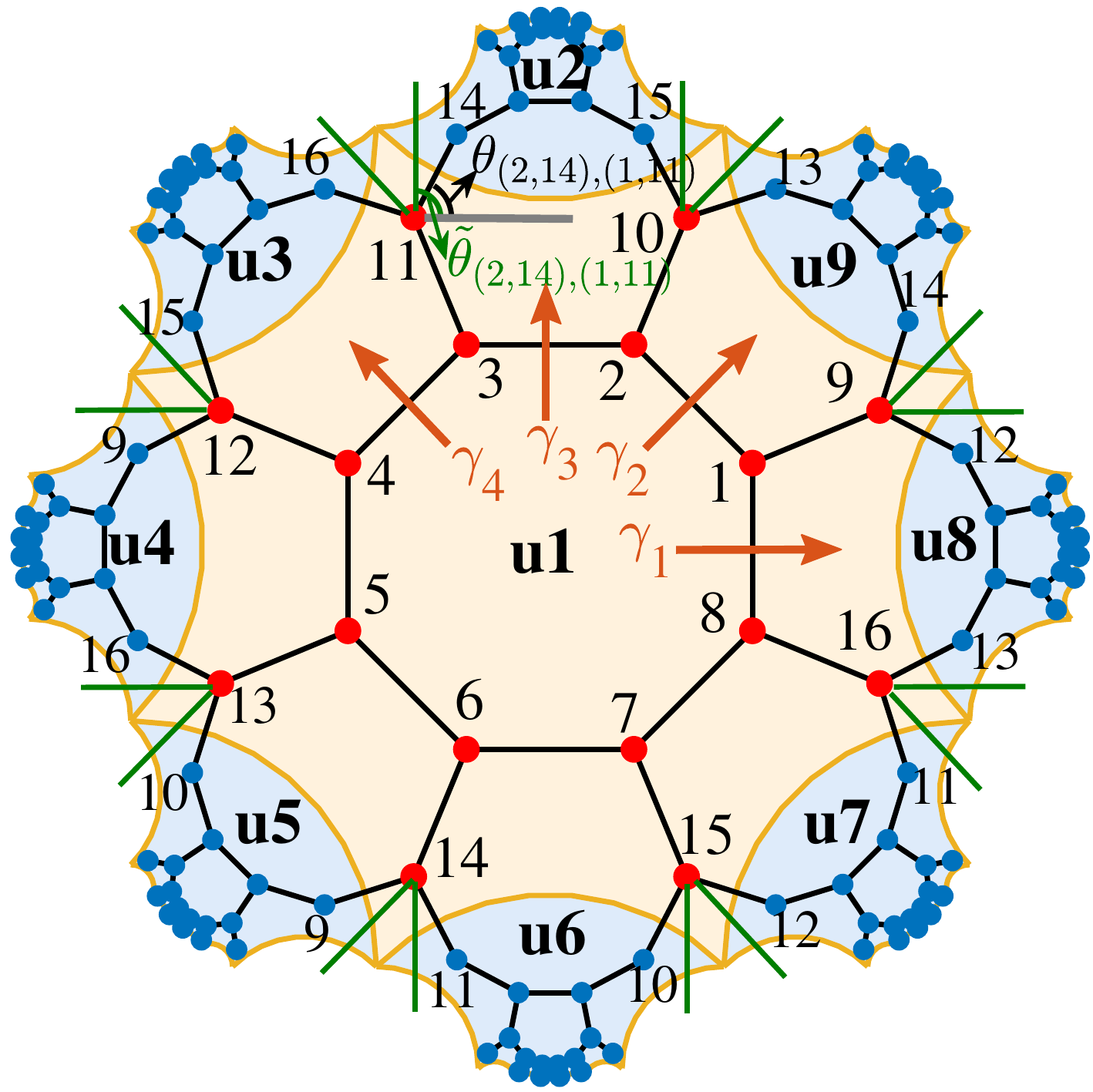}
			\caption{The hyperbolic \{8,3\} lattice with nine unit cells $u1, \dots, u9$. We label each site in a unit cell by numbers. 
				The modified angle for the hopping from the site 11 in the unit cell $u1$ to the site 14 in the unit cell $u2$
				is labeled as $\tilde{\theta}_{(2,14),(1,11)}$, determined by the polar angle of the direction of the generator $\gamma_3$ (green line), in contrast to the angle $\theta_{(2,14),(1,11)}$.  
			}	
			\label{cell}
		\end{figure}
		In this section, we will provide Fig.~\ref{cell} where the modified angle $\tilde{\theta}_{ij}$ is defined and show how to construct the momentum space Hamiltonian via twisted boundary conditions.
		
		Fig.~\ref{cell} displays the hyperbolic \{8,3\} lattice with nine unit cells $u1, \dots, u9$. In the hopping matrix 
		$T(\theta_{ij})$ in the Hamiltonian in Eq. (4) in the main text,  $\theta_{ij}$ is the angle of the vector from the site $j$ to the site $i$.
		Note that here $i$ and $j$ can be either a composite index or a single index uniquely labeling each site. For example, $i=(2,14)$ denotes
		the site 14 in the unit cell $u2$ (see Fig.~\ref{cell}). In Eq. (4) in the main text,
		the hopping from the site 11 in the unit cell $u1$ to the site 14 in the unit cell $u2$ is determined by the angle $\theta_{(2,14),(1,11)}$ labeled in the figure.
		However, this model breaks the translational symmetry. To restore the translational symmetry without breaking the $C_8 T$ symmetry,
		we follow Ref.~\cite{Urwyler2022arxivS} to modify the intercell hopping by replacing $\theta_{ij}$ by a modified one $\tilde{\theta}_{ij}$.
		For example, for the hopping mentioned above, 
		we use the modified angle $\tilde{\theta}_{(2,14),(1,11)}$, the polar angle of the direction of the generator $\gamma_3$ (green line in Fig.~\ref{cell}).
		The other angles for the intercell hopping are also modified similarly (see the green lines in Fig.~\ref{cell}). 
		After that, we apply the translational operations (generated by four generators $\gamma_1,\dots,\gamma_4$)
		to generate the entire Hamiltonian, which preserves both translational symmetry and $C_8 T$ symmetry. 
		
		We now provide the $64\times 64$ momentum space Bloch Hamiltonian based on the hyperbolic band theory~\cite{Rayan2021sciaS,Rayan2022PNASS,Mao2022arxivS}.
		The onsite and hopping term inside the first unit cell (the light orange region in Fig.~\ref{cell}) is
		\begin{equation}
			{H}_0=\sum_{\alpha,\beta}\left[\sum_{i} m |1 i\alpha \rangle [\tau_z\sigma_0]_{\alpha \beta} \langle  1 i \beta| 
			+ \sum_{\left\langle i,j \right\rangle } |1 i \alpha \rangle [T(\theta_{ij})]_{\alpha \beta} \langle 1 j \beta| \right].
		\end{equation}
		The Bloch Hamiltonian is given by
		\begin{equation}
			\label{Hk}
			H({\bf k})=H(k_1,k_2,k_3,k_4)=H_0+ (e^{-\text{i}k_1}H_{8,1}+e^{-\text{i}k_2}H_{9,1}+e^{-\text{i}k_3}H_{2,1}+e^{-\text{i}k_4}H_{3,1} +\text{H.c.}),
		\end{equation}
		where 
		\begin{eqnarray}
			H_{8,1}&=&\sum_{\alpha,\beta} \left[ |1,12,\alpha \rangle [T(\tilde{\theta}_{(8,12),(1,9)})]_{\alpha \beta} \langle 1, 9, \beta|+
			|1,13,\alpha \rangle [T(\tilde{\theta}_{(8,13),(1,16)})]_{\alpha \beta} \langle 1, 16, \beta|
			\right] \\
			H_{9,1}&=&\sum_{\alpha,\beta} \left[ |1,14,\alpha \rangle [T(\tilde{\theta}_{(9,14),(1,9)})]_{\alpha \beta} \langle 1, 9, \beta|+
			|1,13,\alpha \rangle [T(\tilde{\theta}_{(9,13),(1,10)})]_{\alpha \beta} \langle 1, 10, \beta|
			\right] \\
			H_{2,1}&=&\sum_{\alpha,\beta} \left[ |1,15,\alpha \rangle [T(\tilde{\theta}_{(2,15),(1,10)})]_{\alpha \beta} \langle 1, 10, \beta|+
			|1,14,\alpha \rangle [T(\tilde{\theta}_{(2,14),(1,11)})]_{\alpha \beta} \langle 1, 11, \beta|
			\right] \\
			H_{3,1}&=&\sum_{\alpha,\beta} \left[ |1,16,\alpha \rangle [T(\tilde{\theta}_{(3,16),(1,11)})]_{\alpha \beta} \langle 1, 11, \beta|+
			|1,15,\alpha \rangle [T(\tilde{\theta}_{(3,15),(1,12})]_{\alpha \beta} \langle 1, 12, \beta|
			\right],
		\end{eqnarray}
		where $\tilde{\theta}_{(2,i),(1,j)}$ denotes the modified angle from the $j$th site in the unit cell $u1$ to the $i$th site in the unit cell $u2$.
	
	\section{S-3. Proof of the relation on the quadrupole moment}
	In this section, we will prove that $\overline{Q}_{12}=\overline{Q}_{23}=\overline{Q}_{34}=\overline{Q}_{41}$, where $\overline{Q}_{ij}$ is 
	the average quadrupole moment defined in Eq. (2) in the main text. The system in the main text respects the 
	$C_8M_1$ symmetry, and with this symmetry, the Hamiltonian $H({\bf k})$ in momentum space satisfies
	\begin{equation}\label{symH}
		U_{C_8M_1}H(k_1,k_2,k_3,k_4)(U_{C_8M_1})^{-1}=H(-k_4,k_1,k_2,k_3),
	\end{equation}
	where $U_{C_8M_1}$ is a unitary matrix.
	Fixing parameter momenta $(k_3,k_4)$ and spanning square lattice momenta $(k_1,k_2)$, we obtain the set of occupied eigenstates
	$U_{o,12}(k_3,k_4)=\left(|\psi_1 \rangle,|\psi_2 \rangle,\cdots,|\psi_{n_c} \rangle\right)$ with $|\psi_j \rangle$ being the $j$th occupied eigenstate of $H_s$ (one of $n_c=32L^2$ occupied states). One can obtain the column vector $|\psi_j \rangle$ using the eigenstate of $H({\bf k})$, that is, $[|\psi_j \rangle]_{(nm)}=e^{\text{i} {\bf k}_{12}\cdot{\bf r}_n}[|\psi_\lambda({\bf k})\rangle]_m$, where ${\bf k}_{12} \equiv k_1{\bf e}_x+k_2{\bf e}_y$, ${\bf r}_n$ denotes the position of the $n$th unit cell and $|\psi_\lambda({\bf k})\rangle$ is an eigenstate
	of $H({\bf k})$. For clarity, we write down the definition of the quadruple moment as
	\begin{equation}
		Q_{12}(k_3,k_4)=[\frac{1}{2\pi} \mathrm{Im}\log \det ([U_{o,12}(k_3,k_4)]^\dag \hat{D}U_{o,12}(k_3,k_4) )-Q_o ]\ \text{mod}\ 1,
	\end{equation}
	where we have explicitly indicated that $U_{o,12}$ is a function of $k_3$ and $k_4$.
	
	Thanks to the $C_8M_1$ symmetry, $U_{C_8M_1}|\psi_\lambda({\bf k})\rangle$ is an eigenstate of  $H(-k_4,k_1,k_2,k_3)$.
	It follows that applying $U_{C_8M_1}^{\oplus L^2}$ (for an $L\times L$ square lattice) to $U_{o,12}(k_3,k_4)$ leads to $U_{o,23}(-k_4,k_3)U_{\text{exc}}$, that is,
	\begin{equation}
		U_{C_8M_1}^{\oplus L^2}U_{o,12}(k_3,k_4)=U_{o,23}(-k_4,k_3)U_{\text{exc}},
	\end{equation}
	where in the Hamiltonian $H(k_1,q_2,q_3,k_4)$, $k_1$ is replaced by $-k_4$ and $k_4$ is replaced by $k_3$. $U_{\text{exc}}$ realizes the exchange of column
	vectors so as to arrange the occupied eigenstates in a certain order (in fact, the order is not important). 
	As a result, we have
	\begin{align}
		\label{symQ}
		Q_{23}(-k_4,k_3)&=\left\{\frac{1}{2\pi} \mathrm{Im}\log \det [U_{o,23}^\dag(-k_4,k_3) \hat{D}U_{o,23}(-k_4,k_3) ]-Q_o \right\}\ \text{mod}\ 1 \\ \nonumber
		&=\left\{\frac{1}{2\pi} \mathrm{Im}\log \det [U_{\text{exc}} U_{o,12}^\dag(k_3,k_4) (U_{C_8M_1}^{\oplus L^2})^{-1} \hat{D}  U_{C_8M_1}^{\oplus L^2} U_{o,12}(k_3,k_4)  U_{\text{exc}}^{-1}  ]-Q_o \right\}\ \text{mod}\ 1 \\ \nonumber
		&=\left\{\frac{1}{2\pi} \mathrm{Im}\log \det [U_{o,12}^\dag(k_3,k_4) (U_{C_8M_1}^{\oplus L^2})^{-1} \hat{D}  U_{C_8M_1}^{\oplus L^2} U_{o,12}(k_3,k_4)    ]-Q_o \right\}\ \text{mod}\ 1 .
	\end{align}
	Clearly, $\hat{D}$ commutes with the $ U_{C_8M_1}^{\oplus L^2}$, i.e., $\left[\hat{D}, U_{C_8M_1}^{\oplus L^2}\right]=0$. We thus arrive at
	\begin{align}
		\label{symQ2}
		Q_{23}(-k_4,k_3)&=\left\{\frac{1}{2\pi} \mathrm{Im}\log \det [U_{o,12}^\dag(k_3,k_4)\hat{D}U_{o,12}(k_3,k_4)]-Q_o \right\}\ \text{mod}\ 1, \\ \nonumber
		&=Q_{12}(k_3,k_4).
	\end{align}
	Averaging the quadrupole moment over parameter momenta $(k_3,k_4)$, we have the following relations:
	\begin{align}
		\overline{Q}_{12}&=\frac{1}{(2\pi)^2}\int_{0}^{2\pi} dk_3 \int_{0}^{2\pi}dk_4 Q_{12}(k_3,k_4) \\ \nonumber
		&=\frac{1}{(2\pi)^2}\int_{0}^{2\pi} dk_3 \int_{0}^{2\pi}dk_4 Q_{23}(-k_4,k_3) \\ \nonumber
		&=\frac{1}{(2\pi)^2}\int_{0}^{2\pi} dk_1 \int_{0}^{2\pi} dk_4 Q_{23}(k_1,k_4) \\ \nonumber
		&=\overline{Q}_{23}.
	\end{align}
	Similarly, one can also prove that $\overline{Q}_{23}=\overline{Q}_{34}=\overline{Q}_{41}$ and $\overline{Q}_{13}=\overline{Q}_{24}$ by choosing different pairs of parameter momenta.
	
	\section{S-4. The band and topological properties of the semimetal phase in the 4D momentum-space}
	In this section, we will provide more detailed discussion on the band and topological properties of the higher-order topological hyperbolic semimetal phase in 
	the 4D momentum space. In Fig.~\ref{figS1}(a), we plot the gap of $H({\bf k})$ as a function of $g$ in the 4D Brillouin zone for different system sizes. 
	Clearly, the gap of $H({\bf k})$ closes when $g>1.31$, leading to a gapless phase.
	
	To illustrate the gapless structure in the semimetal phase, we plot the gapless region for $g=1.35$ in Fig.~\ref{figS1}(b) with color denoting $k_1$.
	We find eight degenerate nodes. We also display the distribution of the quadrupole moment $Q_{12}$ in the $(k_3,k_4)$ plane in Fig.~\ref{figS1}(c).
	We see that a part in the Brillouin zone exhibits $Q_{12}$ of $0.5$ and the other part has zero $Q_{12}$, resulting in a fractional value of the 
	average quadrupole moment. The degenerate nodes projected on the $(k_3,k_4)$ plane cannot completely separate topologically nontrivial and trivial 
	regimes. This is due to the fact that a quadrupole topological insulator can undergo phase transitions through edge energy gap closing without involving
	any bulk energy gap closing~\cite{Xu2020PPRS}. 
	
	\begin{figure}[htbp]
		\includegraphics[width = 0.9\linewidth]{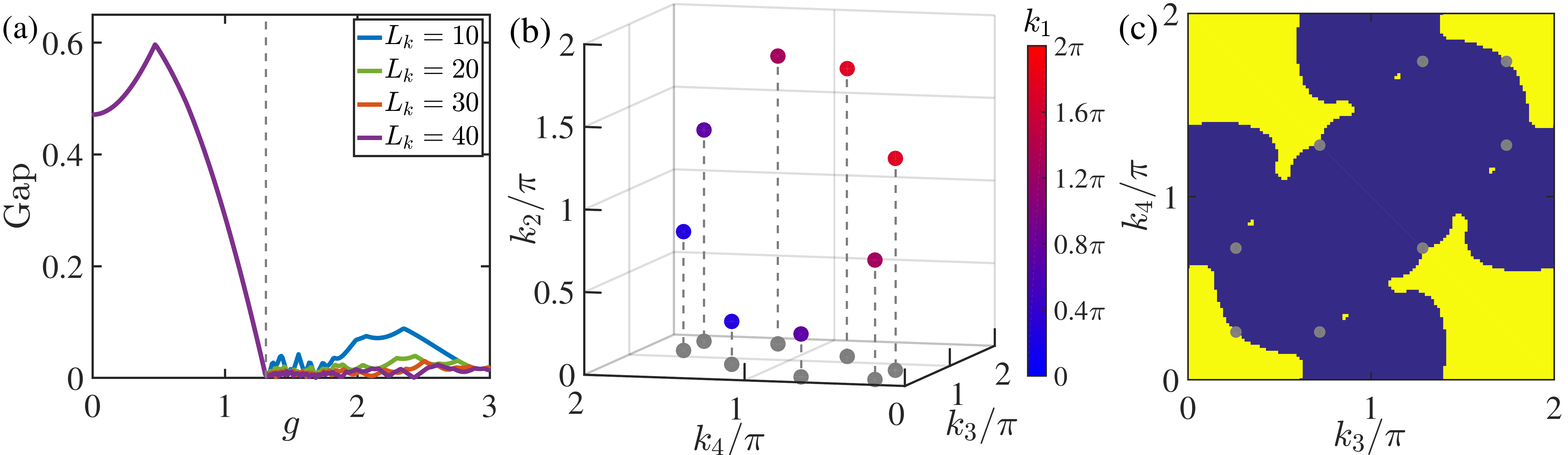}
		\caption{(a) The energy gap of $H({\bf k})$ with respect to $g$ for different system sizes. The 4D Brillouin zone is spanned by $\left\{(k_1,k_2,k_3,k_4)|k_i=2\pi n_i/L_k,\ n_i=0,\dots,L_k-1,\ i=1,\dots,4\right\}$ with $L_k=10,\ 20,\ 30,$ or 40. (b) The degenerate nodes in the 4D energy spectrum of $H({\bf k})$ with $g=1.35$. 
			Color denotes the value of the momentum $k_1$. Gray nodes are the projection of gapless nodes onto the  $(k_3,k_4)$ plane [also see (c)]. 
			(c) The quadrupole moment $Q_{12}$ in the $(k_3,k_4)$ plane for $H({\bf k})$ with $g=1.35$. In the blue region, $Q_{12}=0$, whereas in 
			the yellow region, $Q_{12}=0.5$. Here, $m=0.8$.}	
		\label{figS1}
	\end{figure}
	
		\section{S-5. The energy spectrum and local DOS for the model with translational symmetry}
		In the section, we will show the energy spectrum and local DOS for the model with translational symmetry calculated in a geometry with open boundaries.  
		The energy spectrum in Fig.~\ref{figS-trans}(a) clearly illustrates the existence of zero-energy states, which are localized in the vicinity of a corner position
		as reflected by the local DOS at zero energy in Fig.~\ref{figS-trans}(d). We also see the existence of a gapless phase. 
		However, for larger systems at epoch 5 and 6, we find that the energy spectrum changes dramatically for different system sizes. 
		The consequence is that for a large system, we cannot observe a local DOS with a clear main peak on the boundary [see Fig.~\ref{figS-trans}(f)]. 
		In this case, we can hardly claim that the system is a topological phase in the thermodynamic limit.    
		
		\begin{figure}[htbp]
			\includegraphics[width = 0.8\linewidth]{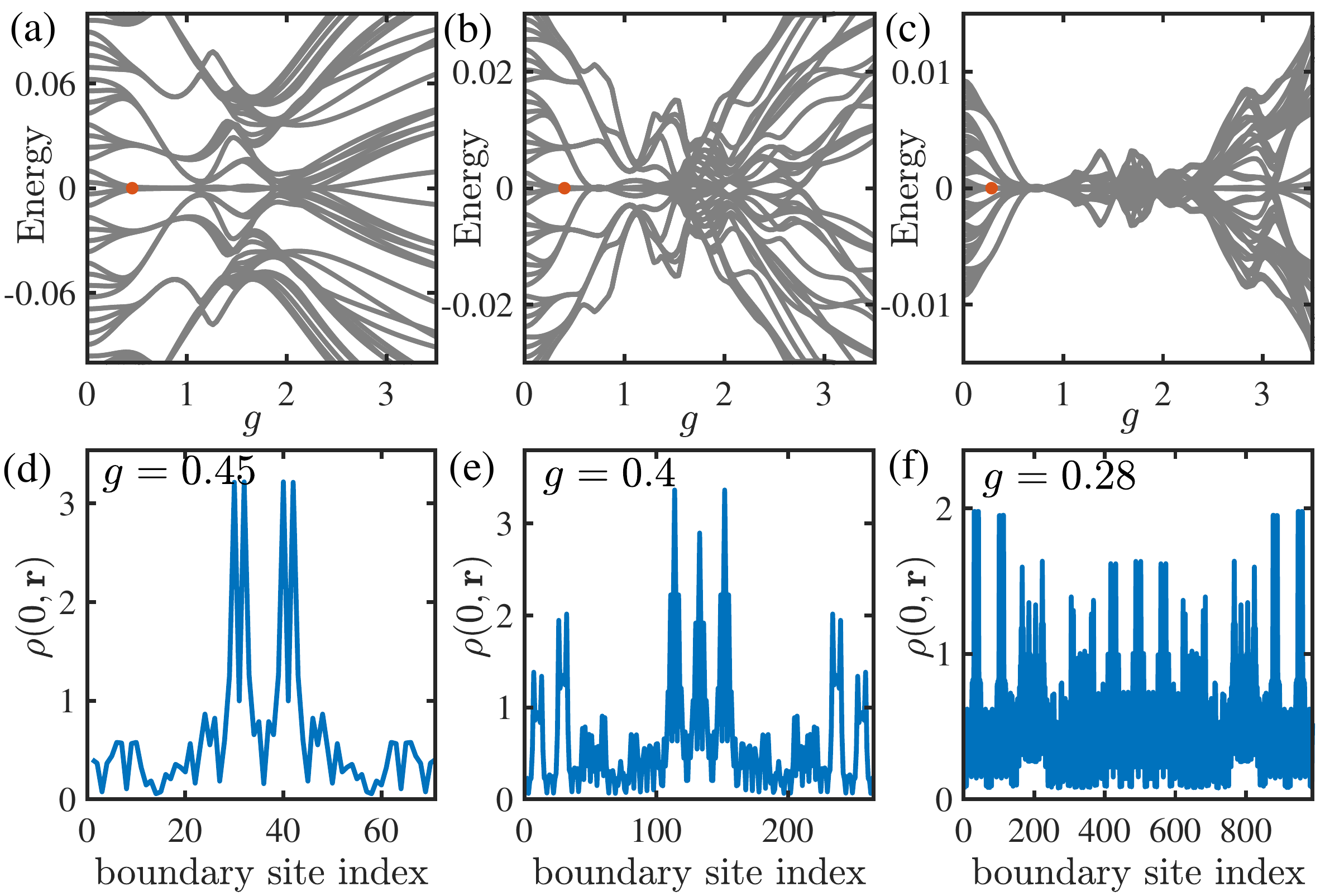}
			\caption{The energy spectrum of the model with translational symmetry versus the system parameter $g$ 
				calculated in real space for a hyperbolic \{8,3\} lattice at (a) epoch 4, (b) epoch 5 and (c) epoch 6.
				The local DOS at zero energy over the $1/8$ sector ($\theta \in [0,\pi/4]$) for (d) $g=0.45$ at epoch 4, (e) $g=0.4$ at epoch 5, and (f) $g=0.28$ at epoch 6.
				The center and two boundary points of the x axis refer to a corner position ($\theta=\pi/8$) and the center of an edge ($\theta=0$ and $\theta=\pi/4$), respectively.
				The corresponding $g's$ are marked out as red circles in (a)-(c).   
				Here, $m=0.8$.
			}	
			\label{figS-trans}
		\end{figure}
	
	\section{S-6. Corner charge}
	For gapped topological phases with corner modes, boundary obstructions lead to the corner-localized fractional charges $\pm e/2$~\cite{Queiroz2021PRBS,Xu2020PPRS}.
	In our case, such corner charges should appear over a 1/8 sector $S_{1/8}$ [see Fig.~\ref{figS3}(a)]. 
	Therefore, we use the net charge over a sector at half filling to characterize the topological feature of the system in real space, which is defined as 
	\begin{align}
		C_S=2N_{S_{1/8}}-\sum_{i\in occ}\sum_{{\bf{r}} \in {S_{1/8}}}\sum_j {|\Psi_{i,j}(\textbf{r})|^2},
	\end{align}
	where $N_{{S_{1/8}}}$ is the number of sites in this sector and $\Psi_{i,j}(\textbf{r})$ is the $j$th component of the $i$th occupied eigenstate at site $\textbf{r}$. 
	The first term arises from  the background positive charge over $S_{1/8}$ (each site
	contributes a $+2$ charge in units of $e$).
	To determine the part contributed by electrons, we introduce a small $\delta$-term $H_{\delta}=\sum_{i,\alpha,\beta}\delta|i\alpha \rangle [\tau_x\sigma_z]_{\alpha \beta} \langle i \beta|$, so that the eightfold degeneracy of the zero-energy states is
	lifted, leading to four corner states with positive energy and the other four with negative energy. As a result, only four corner states are occupied at hall filling. 
	
	In Fig.~\ref{figS3}(b), we plot $C_S$ with respect to $g$. In the deep gapped ($0<g < 1.15$) and reentrant gapped ($1.44 < g < 2$) 
	regimes, $|C_S|$ approaches the quantized value of $0.5$, reflecting that these two regimes are topologically nontrivial. Note that when $g$ is small,
	there is a small gap for the corner modes due to finite-size effects, leading to the inaccurate evaluation of the corner charge. 
	We in fact also plot the corner charge for a larger system at epoch 5, illustrating that it is closer to $0.5$ in the gapped regime
		(see the inset in Fig. 2 in the main text).
	In the gapped trivial phase ($g >2$),
	the corner charge declines to zero. Also note that in the gapless region, the corner charge is not a well defined quantity.
	
	\begin{figure}[htbp]
		\includegraphics[width = 0.6\linewidth]{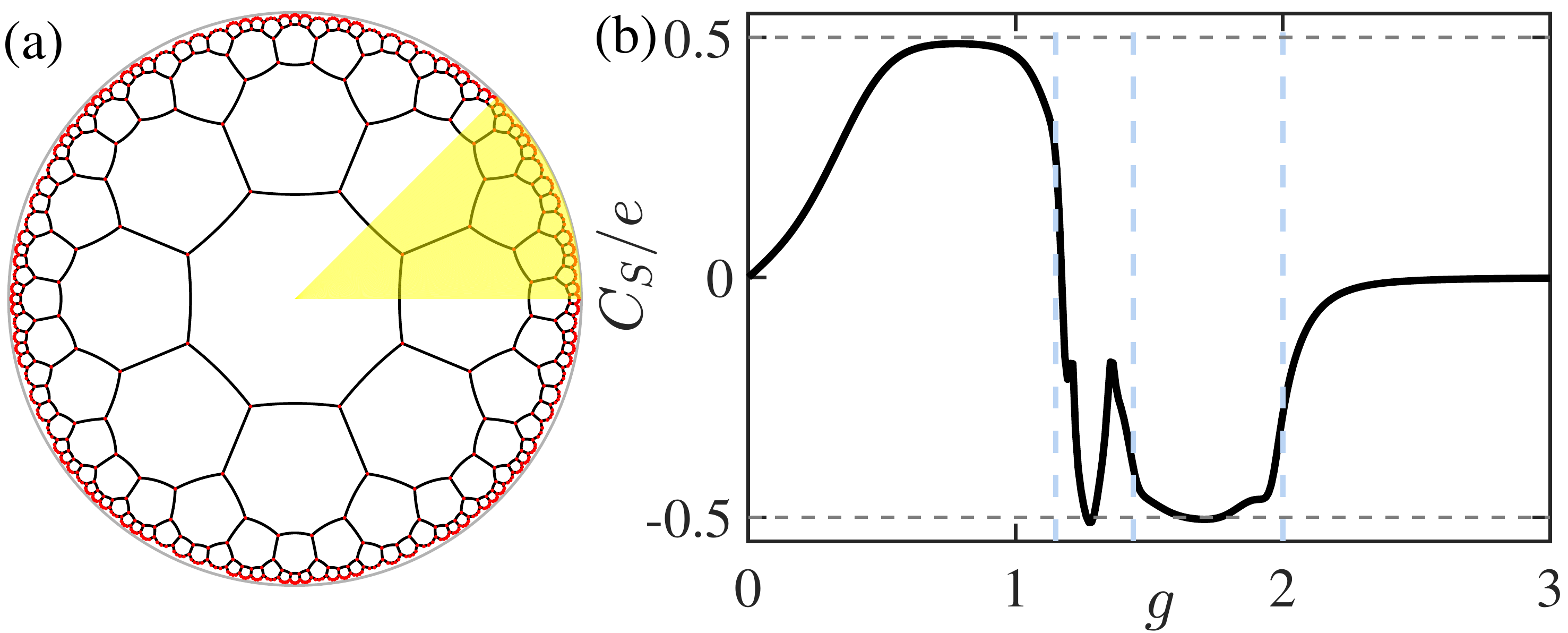}
		\caption{(a) Schematic of the Poincar\'{e} disk model for the hyperbolic \{8,3\} lattice. The yellow region represents the 1/8 sector $S_{1/8}$. 
			(b) $C_S$ with respect to $g$ for the tight-binding model Hamiltonian in Eq. (4) in the main text with 768 sites at epoch 4. Here, $m=0.8$.}	
		\label{figS3}
	\end{figure}
	
		\section{S-7. Finite-size analysis}
		In this section, we will first show that the higher-order topological phase continues to exist in a much larger system even though the mini-gap becomes very small
		and then discuss how the reentrant gapped phase and the branching arise. 
		
		\subsection{A. Energy spectra, energy gap, local DOS for larger systems}
		\begin{figure}[htbp]
			\includegraphics[width = \linewidth]{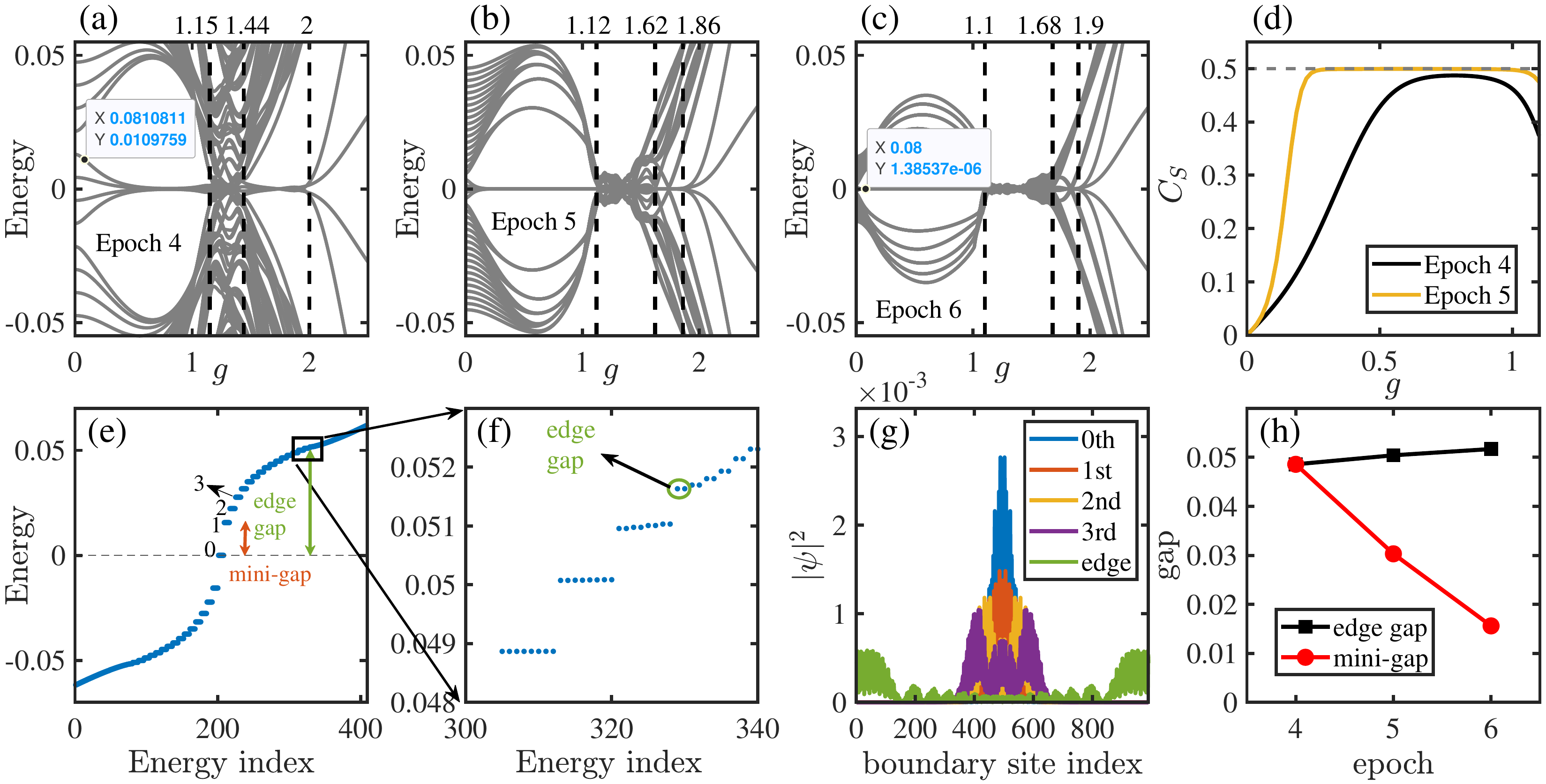}
			\caption{The energy spectrum of the Hamiltonian in Eq. (4) in the main text as a function of $g$ on a hyperbolic \{8,3\} lattice at (a) epoch 4,
				(b) epoch 5 and (c) epoch 6. 
				The vertical lines divide the spectrum into four phases with the transition points labeled above the lines.
				(d) The corner charge with respect to $g$ for the system at epoch 4 (black line) and 5 (yellow line).
				(e) Eigenenergies with respect to energy indices for $g=0.6$ at epoch 6. The mini-gap refers to the first nonzero positive energy and 
				the edge gap refers to the energy where the degeneracy suddenly drops from eight-fold to the double one.
				(f) The zoomed-in view of (e) showing the presence of eight-fold degeneracy at the energy smaller than the edge energy gap.
				(g) The density profiles of the wave functions near zero energy over the $1/8$ sector ($\theta \in [0,\pi/4]$) on the boundary corresponding to the labeled states in (e).
				The green lines describe the density profile of the states encircled by a green circle in (f).
				The center and two boundary points of the x axis refer to a corner position ($\theta=\pi/8$) and the center of an edge ($\theta=0$ and $\theta=\pi/4$), respectively.
				(h) The edge energy gap (black line) and mini-gap (red line) with respect to the epoch number.
				Here, $m=0.8$.
			}	
			\label{figFS1}
		\end{figure}
		
		In this subsection, we will show that the higher-order topological phase may exist in the thermodynamic limit through finite-size analysis. 
		In Fig.~\ref{figFS1}, we plot the energy spectrum for a hyperbolic \{8,3\} lattice at epoch 4, 5 and 6 corresponding to 768, 2888 and 10800 sites, respectively.
		The figure shows that the transition point from the gapped phase to the gapless one changes very slightly with the system size, suggesting 
		its presence in the thermodynamic limit, which is very different from the hyperbolic model with translational symmetry as shown in Fig.~\ref{figS-trans}.
		In this figure, we also see an apparent decline of the energy gap [defined as the mini-gap 
		shown in Fig.~\ref{figFS1}(e)] as the system size increases [also see the red circles in Fig.~\ref{figFS1}(h)]. Despite the decrease of the gap, 
		we are surprised to see that the eight-fold degeneracy of zero-energy modes becomes better. For example, at $g=0.08$, 
		there exists a small energy splitting of $0.01$ for the zero-energy states for a system at epoch 4 [Fig.~\ref{figFS1}(a)];
		however, the splitting drops to about $10^{-6}$ for a much larger system at epoch 6 [Fig.~\ref{figFS1}(c)]. 
		Such a significant decline strongly suggests the existence of eight-fold degeneracy at zero energy and thus the topological phase in the thermodynamic limit.
		To further characterize the topology, we calculate the corner charge $C_S$ for different system sizes and find that it 
		is closer to the quantized value of $0.5$ for a larger system compared with a smaller one [see Fig.~\ref{figFS1}(d)].
		
		Our numerical results also show that the mini-gap states are also eight-fold degenerate [see Fig.~\ref{figFS1}(e)] and are mainly localized near a corner [see Fig.~\ref{figFS1}(g)].
		In fact, we find many eight-fold degenerate states with small finite energies. Interestingly, as we increase the energy, the degeneracy suddenly drops to the double one corresponding
		to the states mainly localized at the center of edges [see the density profile (green line in Fig.~\ref{figFS1}(g)) of the state encircled by the green circle in Fig.~\ref{figFS1}(f)].
		There, the energy spectrum becomes more continuous. We therefore call the gap determined by this energy the edge energy gap 
		[see Fig.~\ref{figFS1}(e)-(f)]. 
		While the mini-gap decreases with the system size, the edge energy gap remains almost unchanged for different system sizes [see Fig.~\ref{figFS1}(h)].
		
		\begin{figure}[htbp]
			\includegraphics[width =0.8 \linewidth]{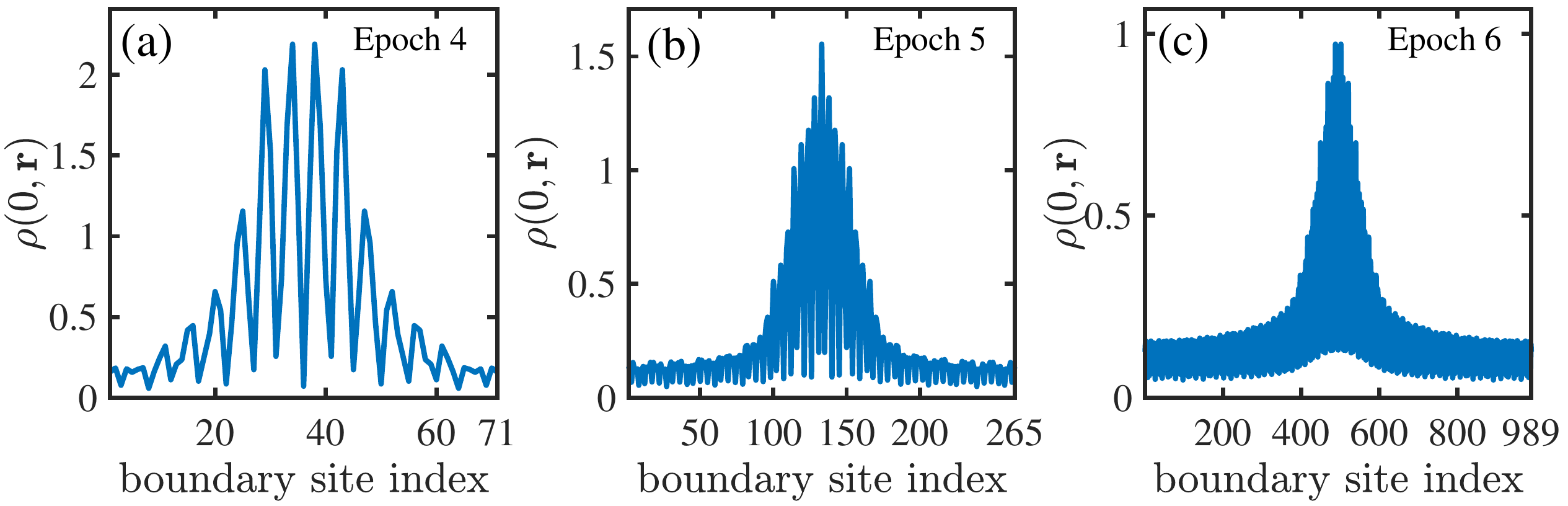}
			\caption{The local DOS at zero energy over the $1/8$ sector ($\theta \in [0,\pi/4]$) on the boundary for 
				the Hamiltonian in Eq. (4) in the main text on a hyperbolic \{8,3\} lattice at (a) epoch 4, (b) epoch 5 and (c) epoch 6.
				The center and two boundary points of the x axis refer to a corner position ($\theta=\pi/8$) and the center of an edge ($\theta=0$ and $\theta=\pi/4$), respectively.
				Here, $m=0.8$ and $g=0.6$.}	
			\label{figFS2}
		\end{figure}
		
		To further confirm the existence of the topological phase in the thermodynamic limit, we plot the distribution of the zero-energy local DOS on the boundary of a hyperbolic lattice 
		at three distinct epochs. All these figures show the existence of peaks of the local DOS near the corner positions. 
		All these results strongly suggest that the topological phase can exist in the thermodynamic limit.
	
	\subsection{B. How do the reentrant gapped phase and the branching arise?}
	In the main text, we show the existence of a reentrant gapped phase. In the section, we will show that it arises from finite-size effects, which 
	open the gap of edge states while leave the gap of corner modes unchanged.
	
	To elaborate on the effect, we calculate the average proportion $p_{\alpha}$ of particles on the boundary for the corner modes ($\alpha=c$) and edge modes ($\alpha=e$)
	by
	\begin{equation}
		p_{\alpha}=\frac{1}{N_{\alpha}}\sum_{i\in S_{\alpha}}\sum_{\textbf{r}\in \mathcal{L}} \sum_j |\Psi_{i,j}(\textbf{r})|^2,
	\end{equation}
	where $\Psi_{i,j}(\textbf{r})$ is the $j$th component of the $i$th eigenstate at site $\textbf{r}$ 
	of the Hamiltonian in Eq. (4) in the main text under open boundary conditions, and 
	$\mathcal{L}$ denotes a set consisting of boundary sites of the hyperbolic lattice. 
	$S_c$ is a set consisting of four eigenstates with lowest positive energy, which correspond to the corner modes in the regime with finite edge energy gap. 
	$S_e$ is a set containing ten eigenstates from the fifth positive energy level to the fourteenth positive level; these modes mainly 
	correspond to the edge modes in the regime with finite edge energy gap. $N_\alpha$ denotes the number of elements in $S_\alpha$.
	
	\begin{figure}[htbp]
		\includegraphics[width = \linewidth]{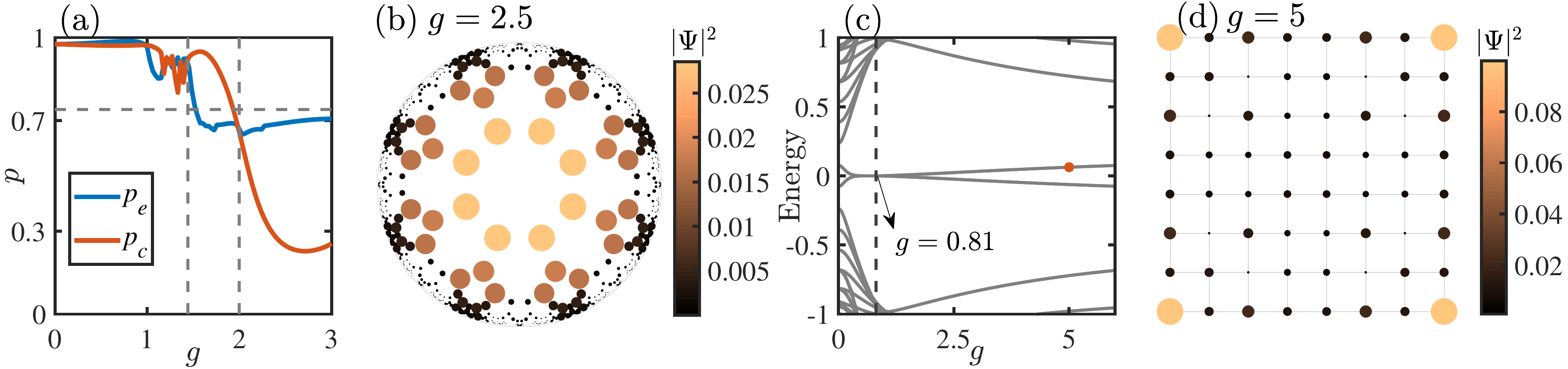}
		\caption{(a) The average proportion $p_\alpha$ of particles on the boundary for the corner modes $\alpha=c$ and edge modes $\alpha=e$ as a function of $g$.
			(b) The density profile of a state nearest to zero energy for $g=2.5$, showing a significant portion of the density permeating into the bulk.
			In (a) and (b), we consider the hyperbolic \{8,3\} lattice at epoch 4 with 768 sites, 
			which is the same as in Fig.~2(b) in the main text, and $m=0.8$.
			(c)
			The energy spectrum of the tight-binding Hamiltonian in Eq.~(4) in the main text as a function of $g$ on a $8\times 8$ square lattice 
			with $m=0.5$ under open boundary conditions.
			(d) The density profile of a state highlighted as red circle in (c) at $g=5$.}	
		\label{figFS3}
	\end{figure}
	
	\begin{figure}[htbp]
		\includegraphics[width = \linewidth]{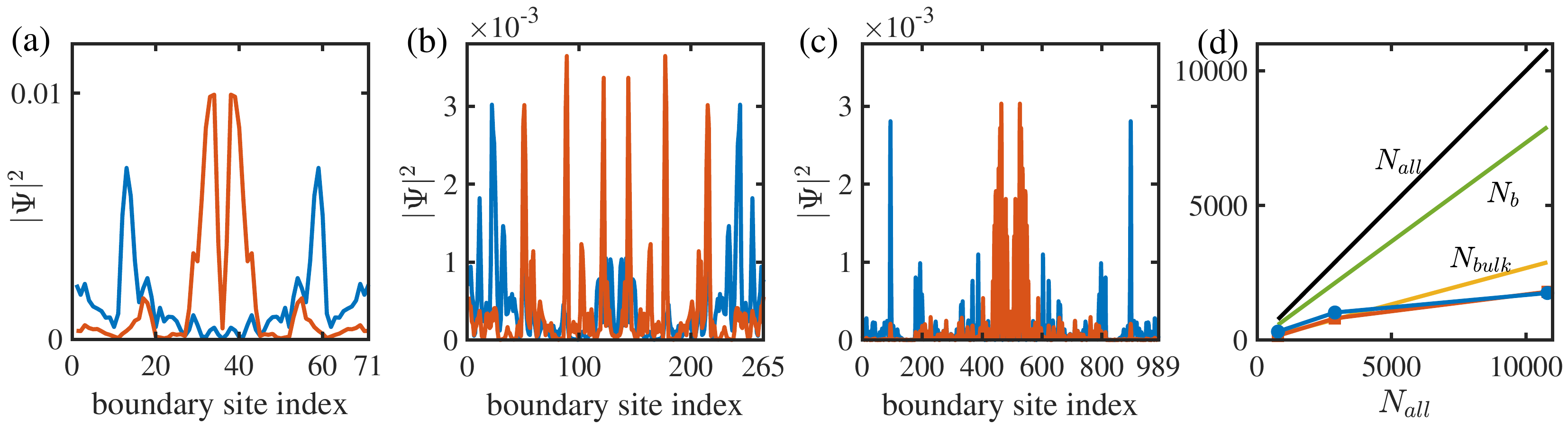}
		\caption{The density profile of eigenstates over the $1/8$ sector ($\theta\in [0,\pi/4]$) on the boundary (in the gapless region) under the lattice structure at (a) epoch 4, (b) epoch 5, and (c) epoch 6. 
			Red (blue) lines denote the states very close to zero energy with negative (positive) Fourier coefficient $b$.
			Here, the center and two boundary points of the $x$ axis refer to a corner position ($\theta=\pi/8$) and the center of an edge ($\theta=0$ and $\theta=\pi/4$), respectively.
			(d) The number of boundary ($N_b$) and bulk ($N_{\text{bulk}}$) sites, and the inverse of the mean IPR with respect to the total number of sites $N_{\text{all}}$. 
			To calculate the mean IPR, we choose 60 eigenstates closest to zero energy forming a set. We then write the set as a union of two sets $S_{-}$ and $S_{+}$, where 
			the density of each state exhibits the negative and positive Fourier coefficient $b$, respectively. 
			The mean IPR is defined as
			$
			I_{\pm}=\frac{1}{N_{\pm}}\sum_{i \in S_{\pm}}\sum_{\bf r}(\sum_{j=1}^4|\Psi_{i,j}({\bf r})|^2)^2
			$,
			where $N_-$ and $N_+$ denote the number of states in $S_{-}$ and $S_{+}$, respectively.
			$1/I_-$ and $1/I_+$ are plotted as red and blue lines, respectively.
			Here, $g=1.3$.}	
		\label{figS-IPR}
	\end{figure}
	
	Figure~\ref{figFS3}(a) shows that as $g$ enters into the reentrant gapped regime, $p_e$ suddenly drops to a value smaller than $N_b/N_{all}$,
	the proportion of boundary sites [$N_b$ is the number of boundary sites and $N_{all}$ is the number of all sites].
	It indicates that the edge states may experience some overlap due to finite-size effects so that their distribution in the bulk increases. Indeed, when 
	we increase the system size, such overlap decreases so that the transition point from the gapless phase to the reentrant gapped one also increases,
	leading to a smaller reentrant regime [see Fig.~\ref{figFS1}(b) and (c)]. We thus expect that in the thermodynamic limit, the phase may disappear.
	The other factor accounting for the presence of the phase is that the corner modes are less sensitive to finite-size effects than the edge states, as reflected by 
	Fig.~\ref{figFS3}(a).
	
		However, when $g$ is further increased to $2$, $p_c$ experiences a significant decline, leading to a finite gap for the corner modes. 
		The decline is also revealed by a 
		significant portion of the wave function permeating into the bulk [see Fig.~\ref{figFS3}(b)]. In fact, such a branching also occurs on square lattices [see
		the energy spectrum of the Hamiltonian in Eq.~(4) in the main text on square lattice in Fig.~\ref{figFS3}(c)]. There, we also see the permeating of the wave function into the bulk.
	
	In the square lattice case, besides the branching phase, we only see the gapped higher-order topological phase. 
	We thus conclude that the specific geometry of hyperbolic lattices not only 
	allows for the existence of higher-order topological phases with the symmetry without crystalline counterpart, but also
	the appearance of several phase transitions with respect to $g$.

	\begin{figure}[htbp]
		\includegraphics[width = 0.7\linewidth]{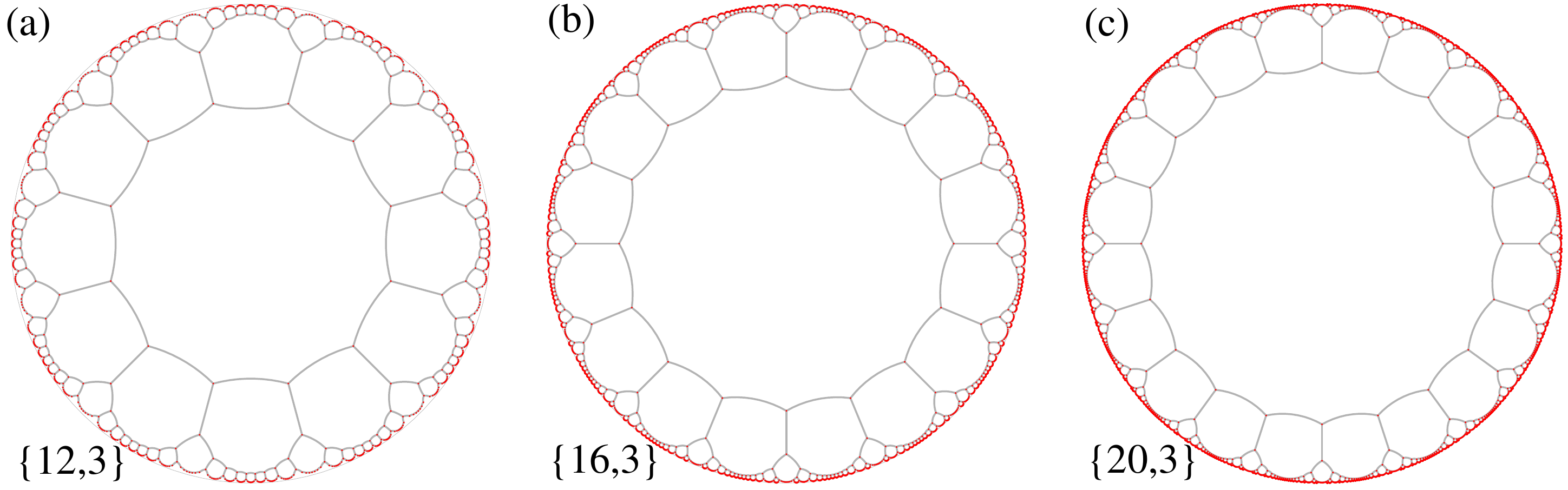}
		\caption{The distribution of vertices at epoch 2 for (a) the hyperbolic \{12,3\} lattice with $\delta=\pi/12$, (b) the hyperbolic \{16,3\} lattice with $\delta=0$, and (c) the hyperbolic \{20,3\} lattice with $\delta=0$.}	
		\label{lattice_more}
	\end{figure}
	
		\section{S-8. Local DOS and IPR in the gapless regime}
		In this section, we provide the density profile of the states very close to zero energy at $g=1.3$ in the gapless phase in Fig.~\ref{figS-IPR}(a)-(c), 
		showing the coexistence of states with large occupation close to a corner or the center of an edge. Interestingly, for the latter states, they have
		well localized property instead of an extended property. To identify the localized property, we plot the number of boundary sites $N_b$ and 
		the number of bulk sites $N_{\text{bulk}}$ as a function of the number of all sites $N_{\text{all}}$, as well as the inverse of the mean inverse participation ratio (IPR)
		in Fig.~\ref{figS-IPR}(d). 
		Different from the crystalline system, in a hyperbolic lattice, $N_b/N_{\text{all}}$ 
		does not change with respect to the epoch number $x$ due to the fact that 
		$N_{\text{all}}(x)\sim e^{\alpha x}$ ($\alpha$ is independent of $x$) and $N_b(x)=N_{\text{all}}(x)-N_{\text{all}}(x-1)$~\cite{Gorshkov2020PRAS,Saa2021arxivS}.
		In the \{8,3\} case, $N_b/N_{\text{all}}=0.73$, and thus the boundary sites take a large proportion.
		Figure~\ref{figS-IPR}(d) shows that the inverse of the IPR for the states near zero energy is much smaller than $N_b$, implying that these states are far from uniformly distributed over the boundary sites.
		Such a localized property may arise from the rapid change of the $\theta_{ij}$ angle between two sites on the boundary, which plays the role of disorder.
		
		\section{S-9. The energy spectrum for the hyperbolic \{12,3\}, \{16,3\} and \{20,3\} lattices }
		In this section, we present the energy spectra of the hyperbolic \{12,3\}, \{16,3\} and \{20,3\} lattices with respect to the system parameter $g$. 
		In Fig.~\ref{lattice_more}, we display their lattice structures at epoch 2. Note that the initial setting about $\delta$ of the central polygon is $\delta=\pi/12$ for the 
		hyperbolic \{12,3\} lattice, and $\delta=0$ for the hyperbolic \{16,3\} and \{20,3\} lattices.

		Figure~\ref{lattice_more_energy} displays the energy spectra of the tight-binding Hamiltonian in Eq. (4) in the main text as a function of $g$ based on these lattice structures. 
		Similar to the \{8,3\} case, there exists a topological region with zero-energy modes (see Fig. 4 in the main text for the local DOS at zero energy for the parameter $g$ marked out as 
		red solid circles). In the \{12,3\} case, we similarly see the presence of a gapless phase. However, the \{16,3\} and \{20,3\} cases do not exhibit the presence of the gapless phase.
		
		\begin{figure}[htbp]
			\includegraphics[width = 0.8\linewidth]{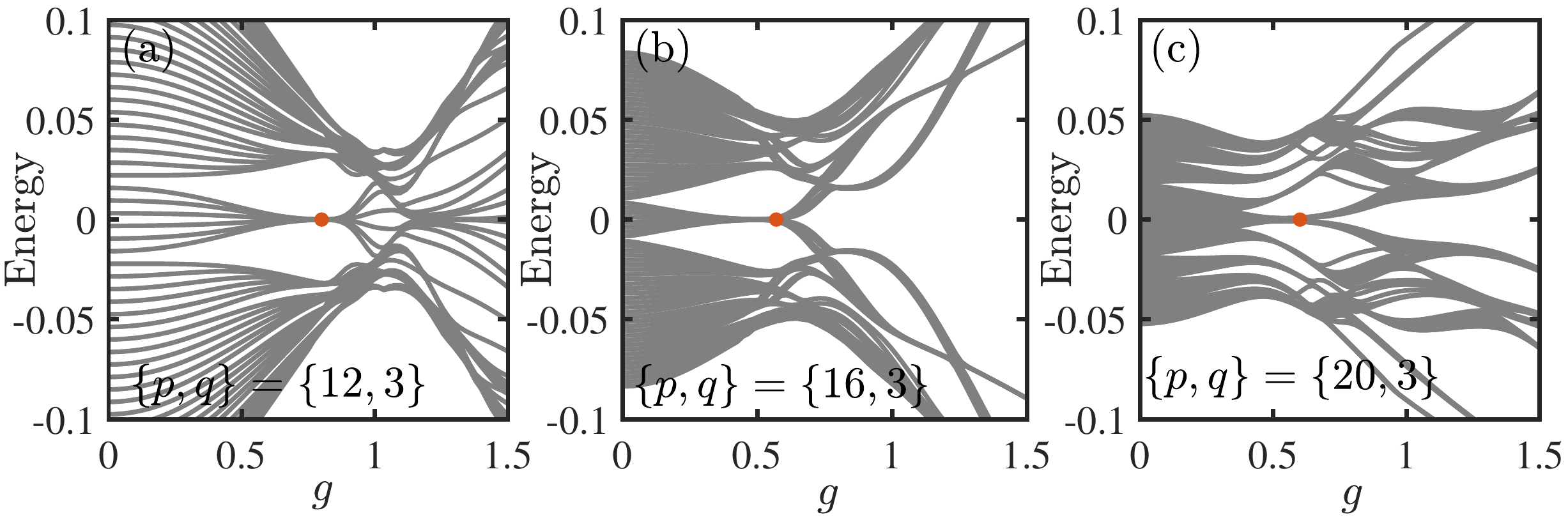}
			\caption{The energy spectrum of the tight-binding Hamiltonian in Eq. (4) in the main text as a function of $g$ for (a) the hyperbolic \{12,3\} lattice with $m=0.9$, (b) 
				the hyperbolic \{16,3\} lattice with $m=0.975$, and (c) the hyperbolic \{20,3\} lattice with $m=1$. Solid red circles highlight the parameters of $g$ used in Fig.~4 in the main text.}	
			\label{lattice_more_energy}
		\end{figure}
	
	\begin{figure}[htbp]
		\includegraphics[width = 0.8\linewidth]{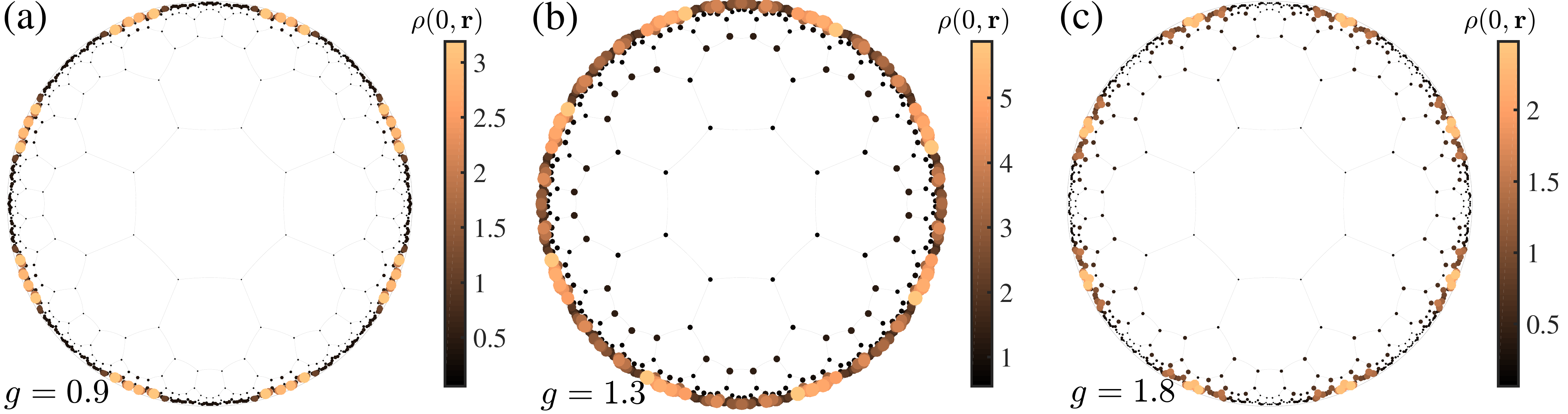}
		\caption{The sample averaged local DOS at zero energy in the presence of weak disorder with $m_r=0.1$ for (a) $g=0.9$,
			(b) $g=1.3$ and (c) $g=1.8$, which correspond to Fig.~3(a-c) in the main text. Here, $m=0.8$.}	
		\label{figS4}
	\end{figure}
	
	\section{S-10. Stability against disorder }
	In this section, we discuss the effect of weak disorder on topological phases by introducing the on-site disorder mass term for the hyperbolic \{8,3\} lattice
	in the Hamiltonian in Eq. (4) in the main text,
	\begin{align}
		\label{Hdis}
		H_{\text{dis}}=\sum_{i,\alpha,\beta}m_r W_i |i\alpha \rangle [\tau_z\sigma_0]_{\alpha \beta} \langle i \beta|,
	\end{align}
	where $W_i$ is a random variable that is uniformly distributed in $[-1,1]$ respecting $C_8$ symmetry and $m_r$ is the strength of disorder. This term respects the chiral symmetry and $C_8T$ symmetry, i.e., $\left[H_{\text{dis}},\Gamma\right]=\left[H_{\text{dis}},C_8T\right]=0$. 
	
	To illustrate that the higher-order topological hyperbolic phases are stable against weak disorder, we calculate the
	local DOS at zero energy averaged over $100$ samples in the presence of weak disorder with $m_r=0.1$.
	Figure~\ref{figS4} shows that the existence of disorder does not destroy the corner modes, 
	indicating the stability of these topological phases against weak disorder.

\end{widetext}


\begin{thebibliography}{99}
\bibitem{footnote} {In Fig.~\ref{Fig2}(a), on the right side of the vertical dashed line, there is a tiny plateau region where the energy gap in momentum space is finite.
	Such a region arises from the edge energy gap closing if we consider open boundaries along the direction of $k_1$ or $k_2$. }
\end{thebibliography}
\end{document}